\documentclass[final,5p,times,twocolumn]{elsarticle}
\usepackage{amsbsy}
\usepackage{amssymb}
\usepackage{amsmath}
\usepackage{enumitem}
\newcommand{\beq}{\begin{equation}}
\newcommand{\eeq}{\end{equation}}
\newcommand{\be}{\begin{eqnarray}}
\newcommand{\ee}{\end{eqnarray}}
\journal{Astroparticle Physics}
\usepackage{graphicx}
\usepackage{slashed}
\usepackage{amsmath}
\usepackage{amssymb}
\usepackage{txfonts}               
\usepackage{longtable,lscape}
\usepackage{hyperref}
\usepackage{tabu}
\newcommand{\bemul}{\begin{multline}}
\newcommand{\eemul}{\end{multline}}
\begin{document}
\begin{frontmatter}
\title{Neutrino production in Population III microquasars}
\author[label1]{Agust\'{\i}n M. Carulli}
\author[label1]{Mat\'{\i}as M. Reynoso}
\author[label2,label3]{Gustavo E. Romero}
\address[label1]{IFIMAR (CONICET-UNMdP) and Departamento de F\'{\i}sica, Facultad de Ciencias Exactas y Naturales, Universidad Nacional de Mar del Plata, Funes 3350, (7600) Mar del Plata, Argentina}
\address[label2]{Instituto Argentino de Radioastronom\'{\i}a (CCT-La Plata, CONICET; CICPBA) \\ C.C. No. 5,  (1894) Villa Elisa, Argentina}
\address[label3]{Facultad de Ciencias Astronómicas y Geof\'{\i}sicas, Universidad Nacional de La Plata \\ Paseo del Bosque s/n, (B1900FWA) La Plata, Argentina.}
%
\begin{abstract}
{Microquasars (MQs) are binary systems composed by a star feeding mass to a compact object through an accretion disk. The compact object, usually a black hole, launches oppositely directed jets which are typically observed in our galaxy through their broadband electromagnetic emission. These jets are considered potential galactic neutrino sources. MQs can also have been formed by the first generations of stars in the universe, i.e., Population III (Pop III) stars, which are considered essential contributors to the ionization processes that took place during the period of “cosmic reionization”.
In the present work, we develop a model that accounts for the main particle processes occurring within Pop III MQ jets, with the aim to obtain the diffuse neutrino flux at the Earth.
We define different zones within the jets of Pop III MQs where particle interactions occur, and primary particles (i.e protons and electrons) are injected. We solve a transport equation for each zone, including the relevant cooling and escape processes, which include $p\gamma$ and $pp$ interactions. Once we obtain the primary particle distributions, we compute the pion and muon distributions, as well as the neutrino output produced by their decays. Finally, we obtain the diffuse neutrino flux by integration over the redshift, the line-of-sight angle, and the MQs lifetime.
We find that, for a range of parameters suitable for Pop III MQ jets, the most relevant site for neutrino production in the jets is the base of the inner conical jet. Additionally, if protons accelerated at the forward shock formed at terminal jet region can escape from the outer shell, they would produce further neutrinos via $p\gamma$ interactions with the cosmic microwave background (CMB). The latter contribution to the diffuse neutrino flux turns out to be dominant in the range $10^7{\rm GeV}\lesssim E_\nu\lesssim 10^9{\rm GeV}$, while the neutrinos produced in the inner jet could only account for a small fraction of the IceCube flux for $E_\nu \sim 10^5$ GeV. The co-produced multiwavelength photon background is also computed and it is checked to be in agreement with observations.}
{}
\end{abstract}
\begin{keyword}
 Radiation mechanisms: non-thermal -- Neutrinos; X-rays: binaries 
\end{keyword}
\end{frontmatter}
%
\section{Introduction}
The first stars formed in the early universe ($z \sim 20-30$) are known as Population III  (Pop III) stars. They were extremely metal poor and with masses within a relative wide range $\sim 10 M_{\odot} - 100 M_{\odot}$ according to star formation simulations \citep{fraser2017}. Their low metallicity and mass imply that nearly $\sim 90\%$ of them collapsed to black holes. Moreover, theoretical results suggest that these stars often formed binary systems ($\sim 50\%$ of them according to Ref. \citep{stacy2013} ). In this context, it is expected that Population III Microquasars (Pop III MQs) can indeed have been generated \citep{sotomayor2018,sotomayor2019}. Composed by a black hole ($\sim 30 M_{\odot}$ in this work) and a Pop III companion star, Pop III MQs {are expected to be super-acreeting systems, i.e., much more powerful than typical galactic MQs \citep{sotomayor2019}. Two of the latter have been detected at high energy gamma rays (Cyg X-1 and Cyg X-3). In addition, the MQ SS433, which is the only super-accreting binary in the Galaxy, was also found to emit very high energy gamma rays \citep{hawcss433}. Given their nature, MQs have long been considered as potential high energy neutrino sources \citep{levinson2001,distefano2002,aharonian2006,production2008,magnetic,zhang2010,anchordoqui2014,possibilities2019}}.

Pop III MQs are also considered as one of the possible main contributors to the process of cosmic reionization \citep{mirabel2011,sotomayor2019}.  
Around $0.37 \ {\rm Myr}$ after the Big Bang, a period known as recombination took place: the plasma of electrons and protons went through a phase transition that coupled them together for the first time to form neutral atomic hydrogen. Therefore, an era named as the ``Dark Ages'' arose, during which no objects capable of producing radiation had yet been formed. Afterwards, the universe passed through a period known as the ``Epoch of Reionization", which began thanks to the ultraviolet (UV) radiation produced by the first formed stars. This radiation might have been capable of ionizing the intergalactic medium (IGM) within the boundaries of the haloes where the stars were born. However, the radiation emanated from this type of sources could not ionize farther away, meaning that another ionization mechanism should have taken place in order to explain the ionization at longer distances. As proposed by {Ref.} \citep{mirabel2011}, X-ray radiation produced by jets arising from accreting BH is essential for the ionization process, since the mean free path of X-rays is much longer than the corresponding to UV radiation. Under these considerations, the reionization capabilities of Pop III MQs jets have also been studied taking into account more complete models \citep{sotomayor2019}. 

Since Pop III MQs would have been formed at high redshifts, gamma rays of very high energy ($E \gtrsim 100 \, {\rm GeV}$) that could have been produced in their jets would have been absorbed during their propagation to Earth. On the other hand, neutrinos would be completely unabsorbed, and hence, it is interesting to compute the neutrino output from these sources in order to assess their  detection with neutrino telescopes such as IceCube. 
{Previous works have addressed the possible neutrino emission arising from other phenomena related to Pop III stars, such as supernova remnants \citep{venya2012,xiao2016} and gamma-ray bursts \citep{schneider2002,gao2011}. Here, we concentrate on a neutrino contribution which has not been previously considered, i.e., the corresponding to the MQ evolutionary phase of binary systems composed by Pop III stars.}

{Specifically}, we develop a model that accounts for the main particle processes occurring within Pop III MQs jets, with the purpose of obtaining a contribution to the diffuse neutrino flux that would arrive on the Earth. To do so, we take into account contributions of neutrino production via the decay of pions and muons resulting from $pp$ collisions between high energy protons and cold proton targets, $p\gamma$ interactions between high energy protons and soft photons produced by electron synchrotron, plus protons interacting with photons from the CMB. We compare the resulting neutrino fluxes with the best fits available for the diffuse flux of astrophysical neutrinos obtained experimentally by IceCube \citep{aartsen2015,aartsen2017}. We also compare the results with the upper limit from the Pierre Auger Observatory \citep{pierreauger2018} and the expected sensitivity for GRAND \citep{GRAND2018}, which would be sensible to  neutrinos of higher energies ($10^7{\rm GeV}\lesssim E_\nu \lesssim 10^{11}{\rm GeV}$). 

This work is organized as follows: in the next section, we describe the jet model and the calculation procedure applied obtain to particle distributions. In the following section, we present the diffuse neutrino flux  for different combinations of parameters, and we also obtain the accompanying flux of multiwavelength photons. Finally, in the last section, we discuss the results and give our concluding remarks.

\section{The model \label{themodel}}
The present model is based on the one presented in Ref. \cite{sotomayor2019}, where the companion star provides the mass that is transferred to the central back hole through an accretion disk in an extremely super-critical regime. {The critical accretion rate of the black hole is associated to the Eddington luminosity $L_{\rm edd}$ ($\dot{M}_{\rm crit} = L_{\rm edd}/c^2$) and thus can be written as $\dot{M}_{\rm crit}=4 \pi GM_{\rm BH}m_p/(\sigma_{\rm T}c)$, where $M_{\rm BH}$ is the mass of the black hole, $m_p$ is the proton mass, and $\sigma_{\rm T}$ is the Thomson scattering cross-section. {The Eddington luminosity is such that for spherically symmetric accretion, the radiation pressure would exactly balance the effect of gravity, implying that higher accretion rates are not attainable. However, if accretion proceeds through a disk as, then it is possible that the emission is directed mostly perpendicularly to the disk plane, thus not cancelling to the accretion process {\citep{ohsuga2007}}. This means that it is perfectly possible to have super-critical systems with accretion rates much higher than the critical one, as it occurs in the mentioned case of SS433. Is is also considered that this regime is appropriate for Pop III MQs due to}} the large amount of mass accreted by the black hole through overflow of the Roche lobe \cite{sotomayor2019}. The MQ phase lasts $\tau_{\rm MQ}\sim 2\times 10^5{\rm yr}$ according to Ref .\cite{inayoshi2017}, and we will take it into account in order to compute the total neutrino emission along the whole life of the MQs. It is also expected that under this regime of accretion, a large fraction of the accreted material has to be ejected in powerful winds and jets that can reach a kinetic power as high as $L_{\rm k}\sim 10^{41}{\rm erg\, s^{-1}}$. Therefore, Pop III MQs are sources expected to be significantly more powerful than their galactic counterparts. Since their formation and existence is well-motivated, we consider that it is worth examining their possible neutrino production potential. For details of the accretion disk, the reader is referred to {Ref.} \citep{sotomayor2019}, while here we concentrate on the mechanisms of neutrino generation, as well as the associated multiwavelength photons resulting from particle acceleration in the jets of Pop III MQs.

  \subsection{The Jets}\label{thejets}

\begin{figure}[htbp] \label{popIIIMQs}
\begin{center}
\includegraphics[width=.47 \textwidth,trim=0 0 0 0,clip]{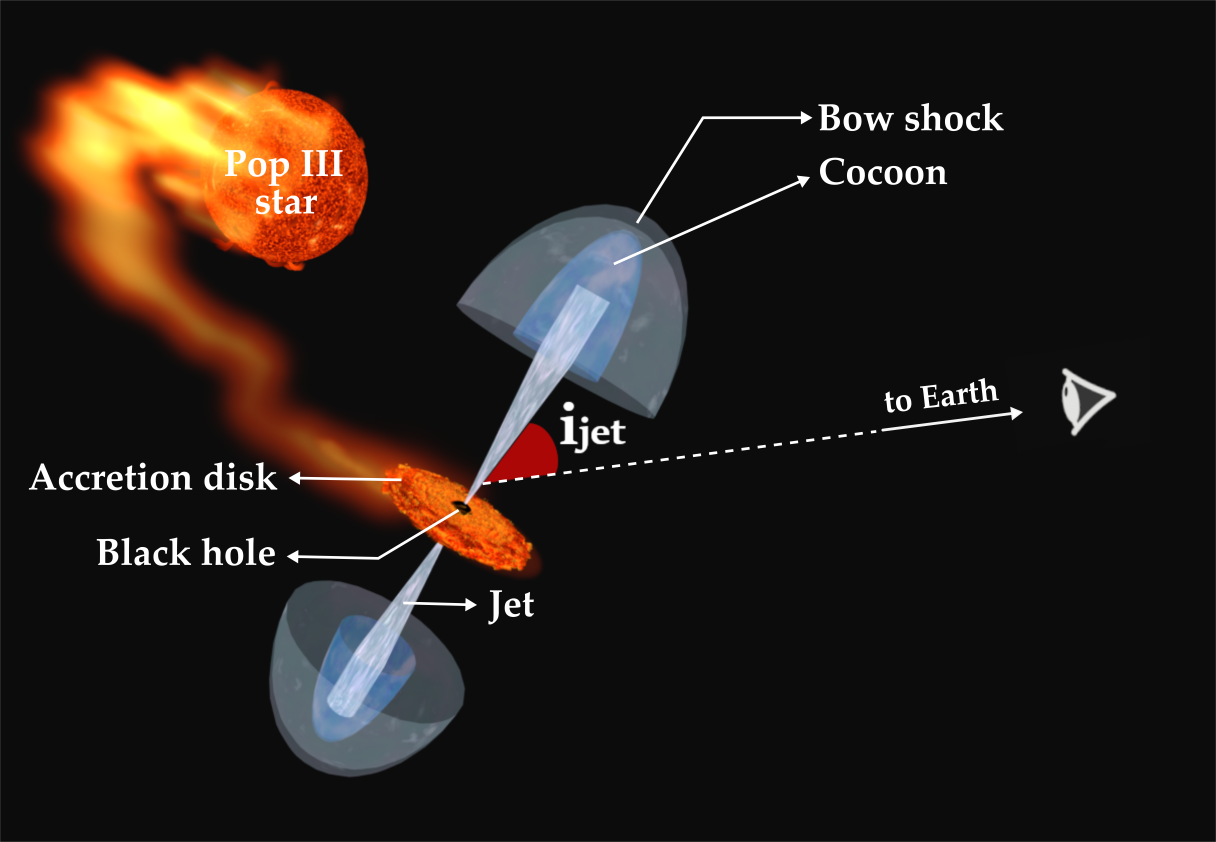}
\end{center}
\caption{{Schematic view of a Population III microquasar. The black hole accretes material from the Population III companion star. The jets arise oppositely directed from the plane containing the accretion disk, and form an angle $i_j$ with the line of sight. Relevant zones, such as the cocoon and the bow shock, are also shown.} \label{fig:popiiimqs}}
\end{figure}

The inner microquasar jets are modeled as two oppositely directed outflows arising from the vicinity of the central black hole (BH) of mass $M_{\rm BH} \simeq 30 M_{\odot}$. {In Fig. \ref{fig:popiiimqs}, we show an schematic view of the main components of the system}. We suppose that each jet is accelerated through the conversion of magnetic to bulk kinetic energy, according to the basic mechanism discussed by {Ref.} \cite{komissarov}. In this scenario, {the flow is initialy magneticaly dominated in the vicinity of the BH and subject to differential collimation decreasing along the jet. It can be shown that this leads to an acceleration of the flow, increasing its Lorentz factor (see Ref. \citep{komissarov} for details).} {Energy} equipartition is expected to hold at an inner position $z_0\sim (10-100)R_g$,  where $R_g = G M_{\rm BH}/c^{2}$ {is the gravitational radius of the accretor}. {For greater distances along the jet ($z_{\rm j}>z_0$), the magnetic energy drops as the jet gradually accelerates, and we assume that the jet reaches its final Lorentz factor $\Gamma$ at a distance $z_{\rm acc}\gg z_0$. At this point, we consider that the jet becomes conical, and thus subject to uniform collimation, which prevents further magnetic acceleration of the flow.}  The magnetic energy at $z_{\rm acc}$ is then supposed to be a small fraction $q_{\rm m}$ of the kinetic energy, i.e., the jet becomes matter-dominated, which is a necessary condition for the development of shocks. Hence, the first region of particle acceleration that we consider is placed at $z_{\rm acc}$ with a size $\Delta z_{\rm b}\sim (1-20)R_{\rm j}${, where $R_{\rm j} = z_{\rm acc} \tan{\psi}$ is the radius of the jet and $\psi$ its half-opening angle}. The jet continues its propagation with a constant velocity, and this is consistent with a magnetic field dependence on the distance along the jet {($z_{\rm j}$)} as {Refs.} \cite{komissarov,vila2008,sotomayor2019}  
\begin{equation}
B(z_{\rm j})=B_{\rm acc}\left(\frac{z_{\rm acc}}{z_{\rm j}}\right), 
\end{equation}
{where $B_{\rm acc}$ is the magnetic field at $z_{\rm acc}$. 
Following Ref. \citep{bordas2009}, we consider further possible zones throughout the jets where some mechanism of particle acceleration can take place \citep{blandford1978,rieger2006,sironi2015}. The inner jets propagate expanding laterally up to the reconfinement point $z_{\rm rec}$, where the pressure in the jets equals that of the external medium. Beyond the reconfinement point, the jets continue their propagation keeping a constant radius until they reach the terminal regions where they are finally stopped by the external medium.

{Since our goal is to obtain the total contribution of neutrinos from Pop III MQs to the diffuse neutrino flux, this is to be found by adding up the individual contributions from the following different zones where emission can take place (see Fig.\ref{fig:zonas} for an schematic view):}
 \begin{itemize}
	\item {Base zone}
	\item {Conical jet}
	\item {Reconfinement zone}
	\item {Cocoon}
	\item {Shell or bow shock}
	\item {External region}
\end{itemize}

 The first zone we consider, where accelerated particles are injected, is close to the base of the jet (base zone, for short), and it is placed in the inner jet at the distance $z_{\rm acc}$ from the BH. There, the kinetic energy density is in sub-partition with the magnetic energy density, i.e. $\rho_{m} = q_{m} \rho_{k}$. {This means that the kinetic energy is smaller by a fraction $q_m$ than it would be under equipartition. The jet kinetic density is:}
\begin{equation} \label{rhokin}
\rho_{k} = \frac{{L_{\rm k}}}{[(\Gamma-1)\Gamma\pi R_{\rm j}^{2} v_{\rm j}]},
\end{equation}
where $L_{\rm k}$ is the jet kinetic power and $v_{\rm j}$ is its bulk velocity. 

\begin{figure*}[tbp] 
\begin{center}
\includegraphics[width=0.7 \textwidth,trim=0 20 0 0,clip]{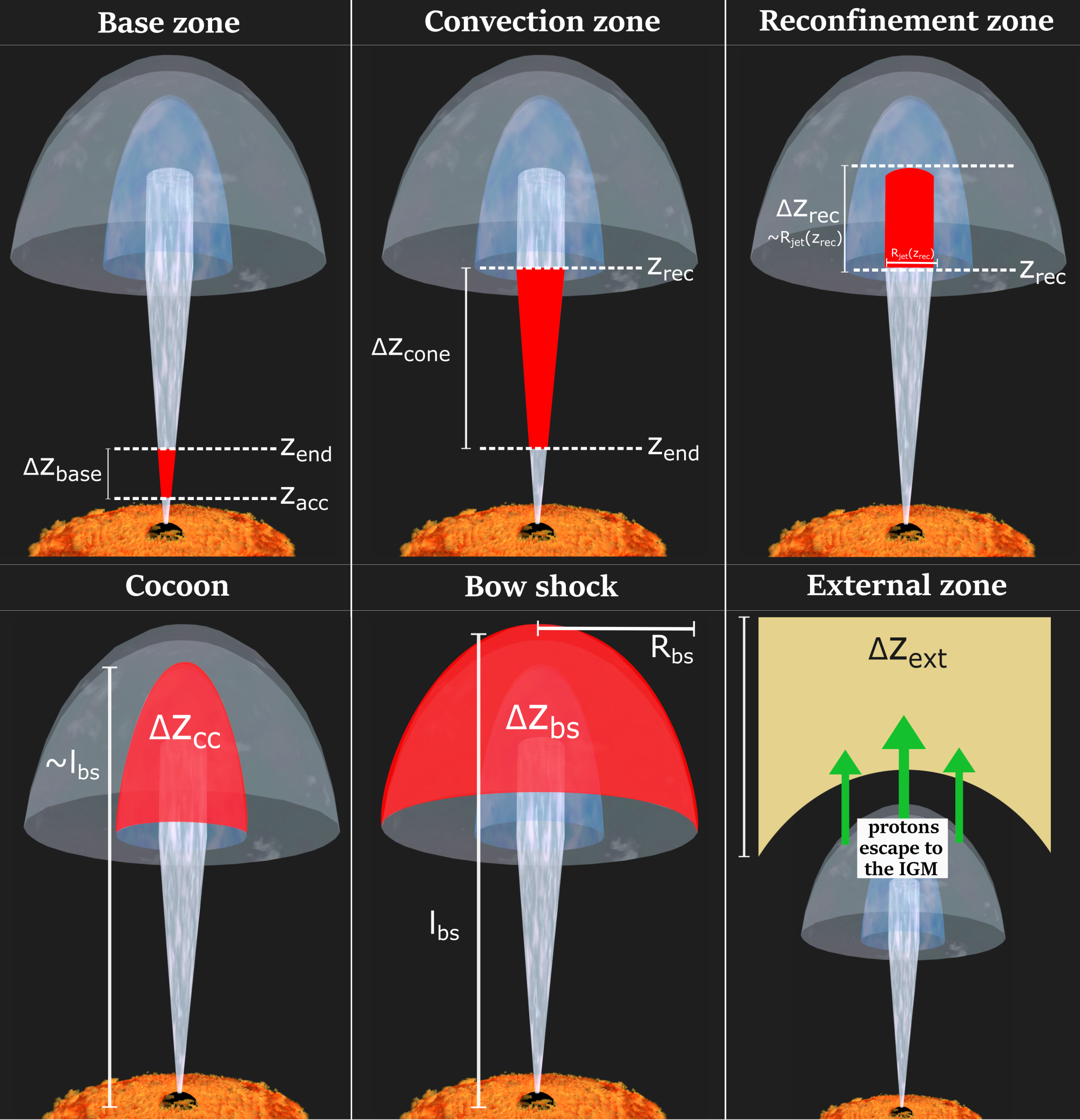}
\end{center}
\caption{Emission zones considered in the jets model. \label{fig:zonas}}
\end{figure*}
  The particles that escape from {the} base zone are injected into a larger conical region at the inner jet, which extends up to the reconfinement point. In order to account for the propagation effect, a convection term is included in the corresponding transport equation, as we discuss below. 
  At the terminal region, the jets interact with the external medium and both a forward shock (bow shock) and a reverse shock are produced. The former propagates into the IGM forming a shell, and a reverse shock that goes inward the jet creating a region called cocoon, which exerts a pressure directed towards the jet. This pressure is able to stop the lateral expansion of the jet and forces the cone-shaped part of the jet to become a cylinder for larger distances. The third emission zone we consider is the reconfinement zone, placed at a distance to the BH that can be computed as {Refs.} \citep{kaiser1997,bordas2009} 
\begin{equation} \label{zrec}
z_{\rm rec} \sim \sqrt{\frac{2L_{\rm k}v_{\rm j}}{\left( \gamma +1\right) \left( \Gamma_{\rm j} - 1\right)  \pi c^{2} P_{\rm coc}}}
\end{equation}
where $\gamma = 5/3$ is the adiabatic index of the cocoon material, $v_{\rm j}$ is the jet velocity, and $P_{\rm coc}$ is the pressure by the cocoon:
\begin{equation} \label{Pcc}
  P_{\rm coc} = \frac{3}{4} \ m_{p} \ n_{\rm IGM}(z) \  v_{\rm bs}^{2}.
\end{equation}
In the latter expression, $n_{\rm IGM}$ is the number density of IGM matter,
\begin{equation} \label{nIGM}
n_{\rm IGM}(z) = \frac{3 H_{0}^{2}}{8 \ \pi \ G \ m_{p}} \  \Omega_{\rm M} \ (1 + z)^{3}
\end{equation}
{where $\Omega_{\rm M}$ is the is the matter density and $H_{0}$ is the Hubble constant, both corresponding to $z=0$.} The velocity of the bow shock is
\begin{equation}
v_{\rm bs} = \frac{3}{5} \frac{ l_{\rm bs}}{t_{\rm MQ}}.
\end{equation}  
In turn, the distance from the BH to the bow shock is given by
\begin{equation}
l_{\rm bs} = \left(\frac{L_{k}}{m_{p} \  n_{\rm IGM}}\right)^{\frac{1}{5}}  t_{\rm MQ}^{\frac{3}{5}} 
\end{equation}
where $t_{\rm MQ}$ is the microquasar age and $L_{\rm k}$ is the jet power.

The remaining two emission zones considered in each MQ jet are placed at the cocoon and the bow shock (shell). However, we still consider that the protons escaping from the shell are injected outside the system in an external zone, where they can further interact with the CMB.

\subsection{Physical processes in the jets}
Here we discuss the main cooling processes that affect the relativistic particles at the zones mentioned above. 
We consider a density of cold protons in each zone of the model given by:
\begin{equation}
n_p = \frac{\rho_{\rm k}}{m_p c^2},
\end{equation}
where $\rho_{\rm k}$ is the kinetic energy density given by Eq.\ref{rhokin}.
These cold protons are considered to be targets for the relativistic protons, and the corresponding rate of $pp$ interactions is \citep{begelman1990}: 
\begin{equation}
t_{pp}^{-1}(E_p) = n_{p} c \sigma_{pp}^{\rm(inel)}(E_p) K_{pp},
\end{equation}   
where the inelasticity coefficient is $K_{pp} \approx 1/2$ since the high energy proton losses half of its total energy per interaction on average, and $\sigma_{pp}^{\rm(inel)}$ is taken to be as in Ref. \citep{kelner2006}.

For each particle type considered (electrons, protons, pions, and muons), the synchrotron cooling rate is,
\begin{equation}
t_{\rm syn}^{-1}(E_i) =   \frac{4}{3} \left(\frac{m_e}{m_i}\right)^3 \frac{\sigma_T B^2}{m_e c \ 8\pi} \frac{E_i}{m_i c^2}, 
\end{equation}
where {$m_e$ is the mass of an electron, and $m_i$ and $E_i$ are the mass and the energy of particles of type $``i"$,  respectively. The particle types considered are electrons ($i=e$), protons ($i=p$), pions ($i=\pi$), and muons ($i=\mu$).} $B$ is the magnetic field corresponding to each of the zones. 

The synchrotron emission of electrons with an energy distribution $N_e$ produces a background density of photons which can be approximated by the following expression in the comoving reference frame:  
\begin{equation} \label{nph}
n_{\rm ph} (E_{\rm ph}) = \frac{\epsilon_{\rm syn}(E_{\rm ph})}{E_{\rm ph}} \frac{R_{\rm j}}{c},
\end{equation}
where {$E_{\rm ph}$ is the photon energy, and} $\epsilon_{\rm syn}$ is the power per unit volume per unit energy of the photons,
\begin{equation} \label{epsilon_ph}
\epsilon_{\rm syn}(E_{\rm ph}) = \left( \frac{1-e^{-\tau_{\rm SSA}(E_{\rm ph}))}}{\tau_{\rm SSA}(E_{\rm ph})} \right) \int_{m_e c^2}^{\infty} dE 4\pi P_{\rm syn} N_e(E).
\end{equation}
{Here, $P_{\rm syn}(E_{\rm ph},E)$ is the power per unit energy of synchrotron photons with energy $E_{\rm ph}$ emitted by an electron of energy $E$. It is defined as  \citep{blumenthalgould1970,rybickilightman2004}:}
\begin{equation}
P_{\rm syn}(E_{\rm ph},E) = \frac{\sqrt{2}e^3 B}{m_e c^2 h} \frac{E_{\rm ph}}{E_{\rm cr}} \int_{\frac{E_{\rm ph}}{E_{\rm cr}}}^{\infty} d\zeta K_{5/3} (\zeta)
\end{equation}

{where $K_{5/3}(\zeta)$ is the modified Bessel function of order $5/3$ and}

\begin{equation}
E_{\rm cr} = \frac{\sqrt{6}heB}{4\pi m_e c} \left(\frac{E}{m_e c^2}\right)^2.
\end{equation}

 {The effect of synchrotron self-absorption (SSA) is taken into account with the factor between parentheses in Eq. (\ref{epsilon_ph}). This corrects the synchrotron emissivity by accounting for the possibility that low energy synchrotron photons may be reabsorbed by electrons \citep{romero2016}. The corresponding optical depth $\tau_{\rm SSA}$ is given by} 
 \begin{equation}
 \tau_{\rm SSA}(E_{\rm ph}) = \int_{m_e c^2}^{\infty} dE \ \alpha_{\rm SSA}(E_{\rm ph}),
 \end{equation}
 {where the SSA coefficient is \citep{rybickilightman2004}:}
\begin{multline}
\alpha_{\rm SSA}(E_{\rm ph}) = \frac{c^2 h^2}{8\pi E_{\rm ph}^3 } \int_{m_e c^2}^{\infty} dE P_{\rm syn}(E_{\rm ph}, E) E^2  \\ \times \left[ \frac{N_e(E-E_{\rm ph})}{(E-E_{\rm ph})^2}  - \frac{N_e(E)}{E^2} \right].
\end{multline}

Inverse Compton (IC) interactions of relativistic electrons with soft photons are considered in the model, in particular, the CMB photons are the most relevant target for the terminal jet zones, while for the electrons in the base zone, the synchrotron emission of the electrons themselves become the dominant target, giving rise to the so-called synchrotron self-Compton (SSC) process. In order to obtain $t_{\rm SSC}^{-1}$, we apply a successive approximation method as we discuss below, since it is necessary to know the particle distribution of the synchrotron emitting electrons: 
\begin{multline}
t_{\rm SSC}^{-1}(E_{e})= \frac{3m_e^2c^4\sigma_{\rm T}}{4E_{e}^3} \int_{E_{\rm ph}^{\rm (min)}}^{E_{e}} dE_{\rm ph} \frac{n_{\rm ph}(E_{\rm ph})}{E_{\rm ph}} \\ \times \int_{E_{\rm ph}}^{\frac{\Gamma_e}{\Gamma_e + 1}E_{e}}dE_\gamma F(q) \left[E_\gamma - E_{\rm ph}\right].
\end{multline}
Here, $E_{\rm ph}^{\rm (min)}$ is the lowest energy of the available background of photons produced by synchrotron of electrons, $q=E_\gamma(\Gamma_e(E_e-E_\gamma))$, with $\Gamma_e=4\,E_{\rm ph}E_e/(m_e^2c^4)$, and the function $F(q)$ is obtained following {Ref.} \citep{blumenthalgould1970}:

\begin{equation}
F(q) = 2 q \ln q + (1+2q)(1-q) + \frac{1}{2} (1-q) \frac{(q\Gamma_e)^2}{1+\Gamma_e q}.
\end{equation}

The cooling of relativistic protons by $p\gamma$ interactions is given by {Ref.} \citep{atoyandermer2003}:
\begin{equation}
t_{p\gamma}^{-1} (\gamma_p) = \int_{\frac{\epsilon_{\rm th}m_pc^2}{2E_p}}^{\infty} d\epsilon \frac{c n_{\rm ph}(\epsilon)m_p^2c^4}{2 E_p^2 \epsilon^2} \int_{\epsilon_{\rm th}}^{\frac{2 \epsilon E_p}{m_pc^2}} d\epsilon_{\rm r} \sigma_{p\gamma}(\epsilon_{\rm r}) K_{p\gamma}(\epsilon_{\rm r}) \epsilon_{\rm r}, \label{tpg}
\end{equation}
where $\epsilon_{\rm th} \approx 150 \,{\rm MeV}$, $\sigma_{p\gamma}$ is the inelastic cross-section for photopion and photopair creation, $K_{p\gamma}$ is the inelasticity coefficient {(taken as in Ref. \citep{atoyandermer2003})}, and $n_{\rm ph} (E)$
represents the density of target photons.

Electrons can also be cooled down by Bremsstrahlung, though it is small compared with the other processes for the parameters considered:
\begin{equation}
t_{\rm Brem}^{-1}(E_e) = 4 \alpha_{\rm FS}  r_e^2  c \, n_p     {\rm ln}\left(\frac{2 E_e}{m_e c^2} - \frac{1}{3}\right).
\end{equation}

The adiabatic cooling rate for a gas of relativistic particles in an expanding volume at a rate $dV/dt$ is \cite{longair2011}:
\begin{eqnarray}
t_{\rm ad}^{-1}(E)= \frac{1}{3V}\frac{dV}{dt}.
\end{eqnarray}
For a conical jet at a distance $z_{\rm j}$ from the central source, considering an element of volume  $V=\pi R_{\rm j}^2\,dz_{\rm j}$ and a lateral expansion velocity as $dR_{\rm j}/dt= v_{\rm j}\,\tan\psi$, yields \citep{boschramon2006}:
\begin{equation}
t_{\rm ad}^{-1} = \frac{2}{3} \frac{\beta c}{z_{\rm j}}.
\end{equation}
{where $\beta$ is the bulk velocity of the jet at that position in units of $c$.}
In the case of spherical expansion (as considered in the bow shock) is:
\begin{equation}
t_{\rm ad}^{-1} =  \frac{\beta c}{\Delta z_j},
\end{equation}
where $\Delta z_j$ is the size of the zone considered.



\section{Relativistic particles at the different zones}
\begin{center}
\begin{table*}
\centering
 \begin{tabular}{|lcccccc|}
\hline
\ \  \footnotesize{Symbol} & \ \ \ \  \footnotesize{Description} & \ \ \ \  \footnotesize{Base} & \ \ \ \  \footnotesize{Reconfinement} & \ \ \ \  \footnotesize{Cocoon} & \ \ \ \  \footnotesize{Shell} & \ \ \ \  \footnotesize{Units}\\ 
\hline
   \ \ \footnotesize{$ L_{p,j}+L_{e,j} $}& \footnotesize{Power injected} & \footnotesize{$ 10^{40}$} &  \footnotesize{$ 10^{39}$} &\footnotesize{$10^{39}$}&\footnotesize{$10^{39}$} &\footnotesize{${\rm erg} \ {\rm s}^{-1} $} \\
    \ \ \footnotesize{$ \Gamma_{j} $} & \footnotesize{Lorentz factor} & \footnotesize{$ {1.25-10} $} & \footnotesize{$1.25-10$} & \footnotesize{$1.25-10$} & \footnotesize{$< 1.003$} & \footnotesize{$1$} \\ 
   \ \ \footnotesize{$ q_{\rm m} $}& \footnotesize{Magnetic parameter} & \footnotesize{$ 5\times10^{-3}$}& \footnotesize{$0.1$ }& \footnotesize{$0.1$}&\footnotesize{ $0.1$ }&\footnotesize{$1$}\\
   \ \ \footnotesize{$ \alpha $}&  \footnotesize{Injection index} & \footnotesize{$1.8-2.2$} & \footnotesize{$1.8-2.2$}& \footnotesize{$1.8-2.2$}& \footnotesize{$1.8-2.2$}& \footnotesize{$1$}\\
    \ \ \footnotesize{$ R_j $} & \footnotesize{Radius of emitter} & \footnotesize{$ 4.4\times 10^{9}-1.8 \times 10^{10}$}& \footnotesize{$1.9\times 10^{19}$}  &   \footnotesize{$4.8\times 10^{20}$} &   \footnotesize{$4.8\times 10^{20}$} &  \footnotesize{$  {\rm cm}  $} \\
   \ \ \footnotesize{$ z_{j} $} & \footnotesize{Injection point} & \footnotesize{$ 4.4\times 10^{10}-1.8 \times 10^{11}$} & \footnotesize{$3.3\times 10^{20}$} & \footnotesize{$2.6\times 10^{21}$}& \footnotesize{$2.6\times 10^{21}$}&\footnotesize{$ {\rm cm} $ }\\
   \ \ \footnotesize{$ B $}& \footnotesize{Magnetic field} & \footnotesize{$ 2 \times 10^4- 9.5 \times 10^4 $} & \footnotesize{$4.2\times 10^{-3}$} & \footnotesize{$3\times 10^{-6}$} &\footnotesize{$4\times 10^{-3}$}& \footnotesize{$ {\rm G} $}\\
       \hline	
       \end{tabular}
\label{tableparamall}
\caption{Main parameters of the model for {the four zones where primary particles are injected as a power-law in the energy. The conical jet and the external zones are not listed because injection there is determined by the escaping particles from the base and from the shell, respectively.} The values for the reconfinement, coccoon, and shell correspond to $z=8$ and $t_{\rm MQ}=6.7\times 10^{4}{\rm yr}$. }
\end{table*}
\end{center}
  
  As mentioned above, populations of relativistic primary particles (electrons and protons) can be accelerated to very high energies by some mechanism such as shock acceleration. At each zone "$j$", the power injected in the form of relativistic particles ($L_{e,j}+L_{p,j}$) is taken to be fraction $q_{\rm rel}$ of the total jet kinetic power ($L_{\rm k}$). We refer to such particles as primary, because these are ones that can initiate the radiation and emission processes that give rise to the production of other, secondary particles which include photons, pions, muons, and neutrinos.  

We consider a steady-state one-zone treatment where the emission region is spatially homogeneous, and the injection and cooling rates are independent {of} time. This approach is applied to the base zone, reconfinement region, cocoon, shell, and also to the external zone considered. Instead, for the extended conical part of the inner jet, we apply an inhomogeneous transport equation which also accounts for the convection effect, as discussed below. 

In the terminal regions {(reconfinement, cocoon, and shell)}, the jet becomes affected by the IGM, which first makes the lateral expansion to cease, and ultimately stops the jet propagation. At the reconfinement point $z_{\rm rec}$ given by Eq. (\ref{zrec}),  recollimation shocks can give rise to further particle acceleration  (see e.g. {Ref.} \citep{bordas2009}). The accelerated particles are injected into a cylinder-shaped zone which extends up to the position of the reverse shock, where the cocoon forms ($z_{\rm j} \simeq l_{\rm bs}$). As mentioned, this zone corresponds to the reconfined jet, which has a size $\Delta z_{\rm cyl}= l_{\rm bs}- z_{\rm rec}$ and a radius $R_{\rm rec}=z_{\rm rec}\tan\psi $, and where additional $pp$, $p\gamma$ interactions can take place. 
Both the bow shock and the reverse shock can generate particle acceleration, and the corresponding emission zones are the shell and the cocoon, respectively. Their radius are $l_{\rm bs}/3$ \citep{bordas2009}, and while the thickness of the shell is $\Delta z_{\rm bs}\simeq R_{\rm bs}$, that of the cocoon is $\Delta z_{\rm coc}=R_{\rm rec}$. In Table \ref{tableparamall}, we present typical values of the main parameters of our model adopted throughout the work. 
  \subsection{Distributions of  particles at the base of the jet, reconfinement region, cocoon, and shell}\label{sec:Reluno}

{In the emission zones placed at the jet base, reconfinement region, cocoon, and shell,} primary particles are injected as a power law in the energy at the comoving frame, for energies greater than $E_{i,\rm min}= 2 m_i c^2$:
  \begin{equation}
  Q_{i,{j}}(E_i) = K_{i,j}E_i^{-\alpha} \mathrm{e}^{({-E_i}/{E_{{\rm max},i,j}})}.
  \end{equation} 
{Here, $``i"$ refers to electrons ($i=e$) and protons ($i=p$), } $\alpha$ is the injection index and $K_{i,j}$ is a constant fixed by normalization on the total power injected in electrons and protons, 
\begin{equation}
L_{i,j}= 4\pi \Delta V_{j,{\rm com}} \int_{E_{i,\rm min}}^\infty dE_i E\, Q_{i,j}(E_i).\label{Eq-normalization}
\end{equation}
This expression is applied in the comoving reference, $E_{i,\rm min}$ is the minimum energy of injection, $\Delta V_{j,\rm com}=\Gamma\Delta V_j$ is the comoving volume of the corresponding zone, and $\Delta V_j$ is the Lorentz contracted volume as seen from the BH frame.
 %
%
The maximum energies $E_{{\rm max},i,j}$, which appear in the exponential cut-off, correspond, in principle, to the balance energies for which the total rate of cooling plus escape is equal to the acceleration rate. {The latter is obtained by applying the general requirement that the timescale for energy gain is greater than $r_{\rm gyr}/c$, where $r_{\rm gyr}= E_i/(e\,B)$ is the gyroradius for a particle with energy $E_i$ and charge $e$. Therefore, the acceleration rate is expressed as }
\begin{equation}
t_{\rm acc}^{-1}(E_i)=\eta \frac{e c B}{E_i},
\end{equation} 
{where $\eta<1$ is an efficiency coefficient that depends on the details of the acceleration mechanism \citep{begelman1990}}. One further requirement is} that the particles can only remain confined inside the zone if their gyroradius does not exceed the size of the acceleration region. Thus, it must be fulfilled that $E_i/eB(z_{\rm j}) < R_{\rm j}(z_{\rm j})$, which is known as Hillas criterion, $E_{\rm H}= eB(z_{\rm j})R_{\rm j}(z_{\rm j})$. Therefore, if the balance energy $E_{{\rm max},i}$ mentioned above happens to be higher than $E_{\rm H}$, then we simply set $E_{{\rm max},i}= E_{\rm H}$.
We show in Fig. \ref{fig:pecool} the proton and electron cooling rates at the base zone and at the shell obtained for redshift $z=8$ and $t_{\rm MQ} \sim 6.7 \times 10^4\,{yr}$. We choose as representative values
 $\Delta z_{\rm j}=5 R_{\rm j}$, $\alpha= 2$,  $\Gamma=1.67$, and $q_{\rm m}=5\times 10^{-3}$, which are included in the ranges indicated in Table \ref{tableparamall}. 
 {We also show in the figure two different cases for the escape rates. One is determined by a constant escape timescale in the comoving frame, $T_{\rm esc}\simeq \Gamma_j\Delta z/v_{j}$, where $\Gamma_{j}$ is the Lorentz factor of the zone considered and $v_{j}$ is its velocity. The other case considered corresponds to a Bohm diffusion timescale, $T_{\rm B}(E)= {(\Delta z)^2}/{[2\,D_{\rm B}(E_i)]}$, where the diffusion coefficient is $D_{\rm B}(E_i)=r_{\rm gyr}c/3$, so that  }
$$
T_{B}(E_i)=\frac{3\,e\,B(\Delta z)^2}{2\,E_i c}.
$$


\begin{figure*}[htbp] \label{pecool}
\begin{center}
\includegraphics[width=1 \textwidth,trim=0 10 99 0,clip]{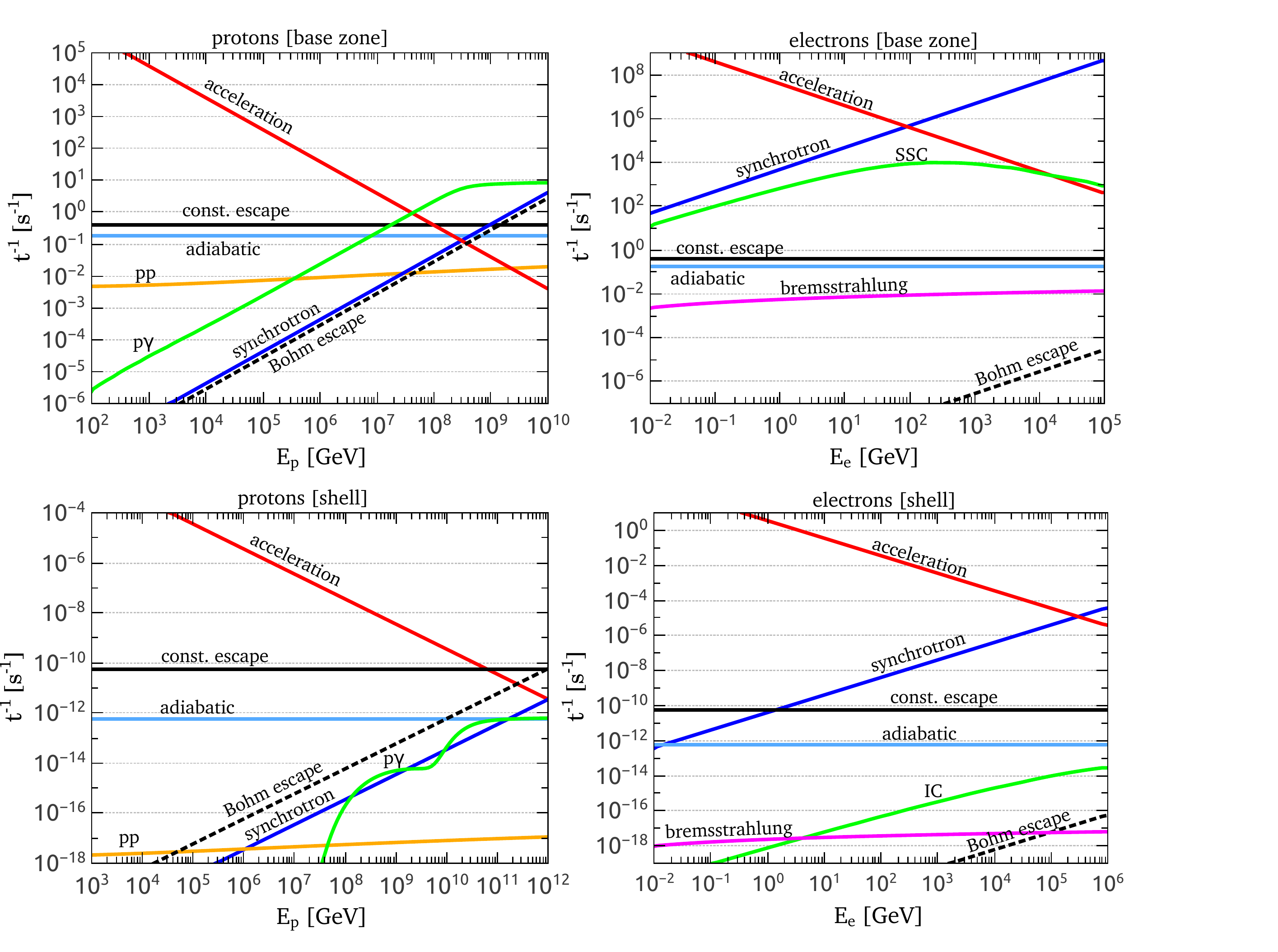}
\end{center}
\caption{Proton (electron) cooling rates for the base zone and bow shock are shown on the left (right) panels. The bow shock rates correspond to a redshift $z = 8$ and $t_{\rm MQ} \sim 6.7\times 10^4{\rm yr}$ \label{fig:pecool}. We adopt $\Delta z_{\rm j}=5 R_{\rm j}$, $\alpha= 2$,  $\Gamma=1.67$, $\eta=0.1$, and $q_{\rm m}=5\times 10^{-3}$.}
\end{figure*}

 
{We then calculate the distributions of primary particles $N_{i,j}$ in the zone ``$j$" by solving the steady-state transport equation:}
 \begin{equation} \label{TransportEQ}
 \frac{d\left[b_{i} N_{i,{j}}(E_i)\right]}{dE} + \frac{N_{i,j}(E_i)}{T_{\rm esc}} = Q_{i,j}(E_i),
\end{equation}
{where $i=e$ stands for electrons and $i=p$ for protons.} $b_{i} \equiv dE_i/dt= -E_i t^{-1}_{\rm cool}$ embody the continuous energy losses of the particles due to the cooling processes that occur in the zone, i.e.{,} synchrotron, IC (or SSC), adiabatic expansion, $pp$ and $p\gamma$ interactions.
In the case of the base zone, we perform successive approximations in order to obtain the electron distribution $N_{e,{\rm b}}$, since the SSC cooling rate {cannot} be neglected. First, we obtain $N_{e,{\rm b}}^{(0)}$ as a solution of the transport equation without considering SSC interactions. Then, we calculate the SSC cooling rate with the obtained $N_{e,{\rm  b}}^{(0)}$, and then we include it in the transport equation to obtain a new approximation $N_{e,{\rm b}}^{(1)}$. We iterate this process until it converges to the correct $N_{e,{\rm b}}$. 

\begin{figure*}[htbp] \label{Nprimary}
\begin{center}
\includegraphics[width=1 \textwidth,trim=17 550 12 0,clip]{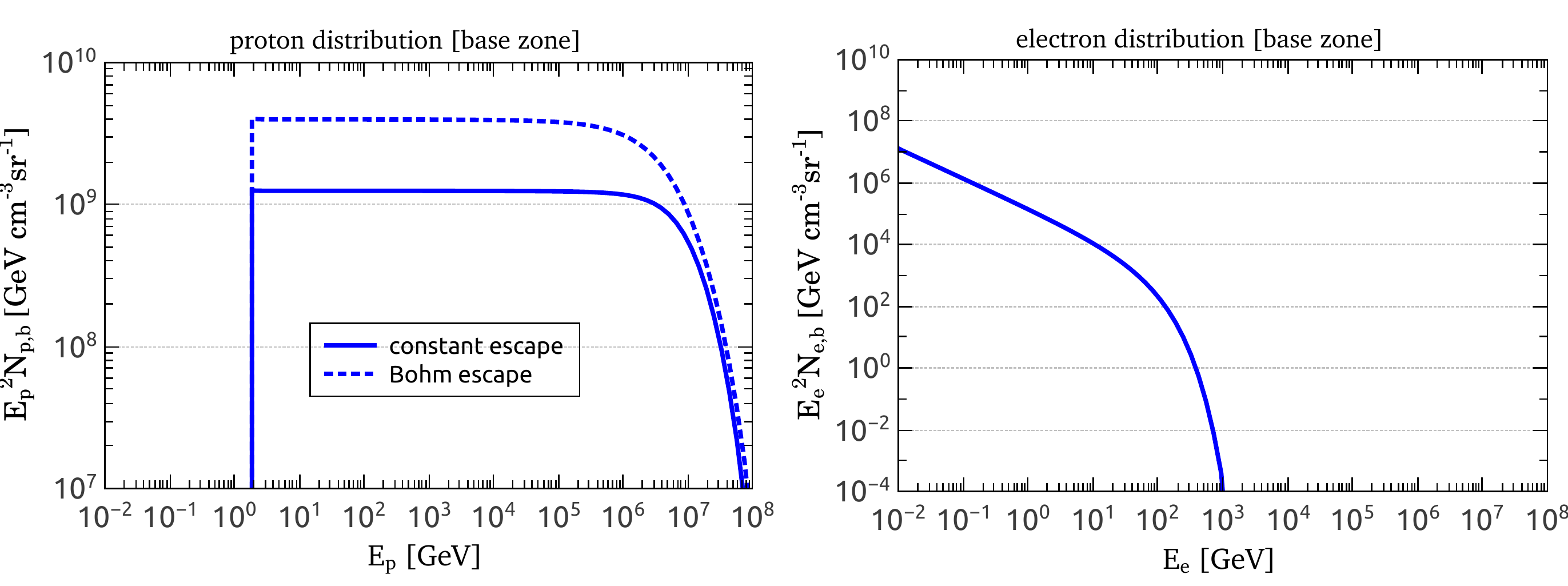}
\end{center}
\caption{Primary particle distributions as a function of the energy at the base zone {in the cases of a constant escape rate (solid lines) and Bohm escape rate (dashed lines)}. \label{fig:Nprimary}}
\end{figure*}

The solution of Eq. (\ref{TransportEQ}) is given by:
\begin{equation}
N_{i,j}(E_i)= \int_{E_i}^{\infty} dE' \frac{Q_{i,j}(E')}{|b_i(E')|} \exp{\left[-\int_{E_i}^{E'}\frac{dE''}{T_{\rm esc} |b_i(E'')|}\right]},\label{TransportEQsol}
\end{equation} 
and
we show in Fig. \ref{fig:Nprimary} the distributions $N_{e,{\rm b}}$ and $N_{p,{\rm b}}$ obtained for the base zone using with the same parameter values as for Fig. \ref{fig:pecool}. {It can be seen in the left panel of Fig.\ref{fig:Nprimary} that the proton distribution is higher in the case of Bohm escape as compared to the faster constant escape case. The dependence with the energy is still the same, since the dominant cooling process in the Bohm escape case is adiabatic cooling, which is constant. In the case of electrons (right panel of Fig. \ref{Nprimary}), there is no difference between the Bohm and the constant escape cases since synchrotron emission largely dominates.} 


Once we obtain the solution of Eq. (\ref{TransportEQ}), it is possible to compute the injection of charged pions produced by $p\gamma$ and $pp$ interactions. We compute the pion injection using accurate approximations to the SOPHIA code for $p\gamma$ interactions \cite{sophia2000} given in Ref. \cite{hummer2010}:
\begin{equation}
Q_{p\gamma\rightarrow\pi^\pm}(E)= \int_{E_p}^\infty \frac{dE_p}{E_p}N_{p,\rm b}(E_p)\int_{\epsilon_{\rm th}}^\infty n_{\rm ph}(E_{\rm ph})R_{\pi}(E_p,E_{\rm ph}),
\end{equation}
where $R_\pi(E,E_{\rm ph})$ is a function that depends on the cross section $\sigma_{p\gamma}$ and includes the different channels for pion production, as discussed in Ref. \cite{hummer2010}. 

\begin{figure*}[htbp] 
\begin{center}
\includegraphics[width=1 \textwidth,trim=7 552 24 0,clip]{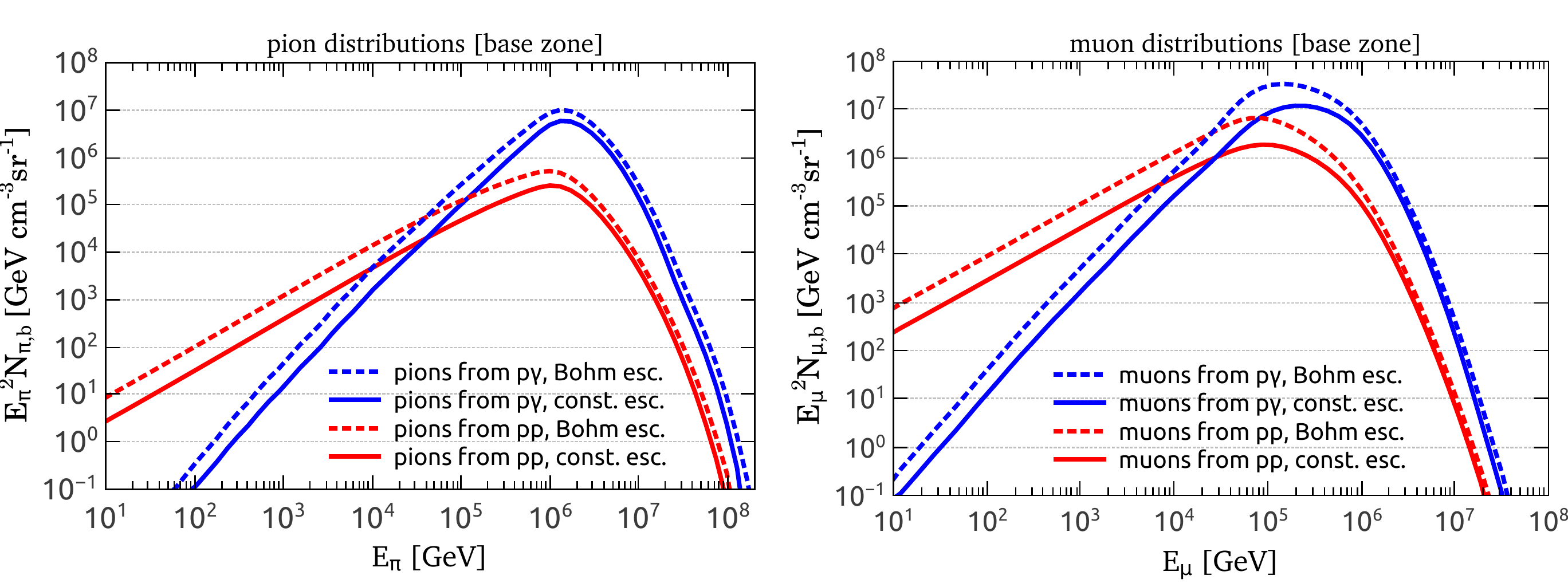}
\end{center}
\caption{Distributions of secondary pions and muons as a function of energy at the base zone {in the cases of a constant escape rate (solid lines) and Bohm escape rate (dashed lines).} \label{fig:Nsecondary}}
\end{figure*}

 As for the injection due to $pp$ interactions, we compute it as
\begin{equation}
Q_{pp\rightarrow\pi^\pm}(E)= n_p c \int_{E}^\infty N_{p,\rm b}(E_p)F_\pi(E_p,E)\sigma_{pp}(E_p),
\end{equation} 
 where $\sigma_{pp}(E_p)$ is the $pp$ cross section and $F_\pi$ is a fitting function given in Ref.\citep{kelner2006} to reproduce the outputs of the SIBYLL simulation code \citep{sibyll1994}.
In order to obtain the pion distribution, we include the corresponding decay term in the trasport equation:
 \begin{equation}\label{TransportEQdec} 
 \frac{d\left[b_{i} N_{i,j}(E_{i})\right]}{dE_{i}} + \frac{N_{i,j}(E_{i})}{T_{\rm esc}} + \frac{N_{i,j}(E_{i})}{T_{i,\rm d}(E_{i})}= Q_{i,j}(E_{i}),
\end{equation}
where $T_{i,\rm d}$ is the particle {lifetime} {and $``i"$ refers to pions ($i=\pi$) and muons ($i=\mu$)}. Once $N_{\pi,j}$ is obtained, we can compute the injection of the muons generated by pion decays $Q_{\mu,j}$ applying the formulae given in Ref. \cite{lipari2007}, which  account for the kinematics of the decay process.
We then plug the muon injection into Eq. (\ref{TransportEQdec}) and compute the corresponding muon distribution $N_{\mu,j}$ using an analogous expression to Eq. (\ref{TransportEQsol}). The result is shown in Fig. \ref{fig:Nsecondary}, along with the pion distribution $N_{\pi,j}$, both corresponding to the base zone with the same parameter values as in the previous figures.

\subsection{Distributions of relativistic particles along the conical jet}

The electrons injected at the base zone suffer severe synchrotron losses for the values
of jet power and magnetic field considered, i.e., the radiative cooling rate dominates over the escape rate. Therefore, the power injected by the electrons at the base zone is completely radiated there before the electrons escape. Conversely, as it can be seen in Fig. \ref{fig:pecool}, the escape rate of protons dominates, along with the adiabatic cooling rate, which means that protons do not participate very efficiently in the cooling processes at the jet base. Thus, a significant fraction of them escape from the base zone to continue their propagation along the rest of the cone-shaped inner jet. The position in the jet where the {conical jet} zone begins is $z_{\rm eoi}=z_{\rm acc}+\Delta z_{\rm b}$, which is the end of the base zone. In turn, the end of the extended conical region is determined by the reconfinement point $z_{\rm rec}$, where, as mentioned above, the pressure exerted by the cocoon changes the geometry of the jet into a cylinder. Supposing that the total rate of {protons escaping from the base zone} is equal to the total rate of protons injected in the conical jet zone,
$$
 N_{p,{\rm b}}(E_p) t_{\rm esc}^{-1}\Gamma\Delta V_{\rm b}= \int Q_{p,{\rm c}}(E_p) dV_{\rm c}, 
$$
it follows that the injection term in the second zone can be expressed by:
\begin{equation}
Q_{p,{\rm c}}(z_{\rm j},E_p) = \frac{\Delta V_{\rm b}}{\Gamma\pi R_{\rm j}^{2}(z_{\rm eoi})} N_{p,\rm b}(E_p) t_{p,\rm esc}^{-1} \delta(z_{\rm j} - z_{\rm eoi}). \label{Qpcone}
\end{equation}

In order to obtain the distribution of protons along the extended conical jet region, we consider a more general transport equation with a convection term. {It is convenient to expresses it using spherical coordinates, so that the transport equation reads \citep{zdziarski2014}:}
\begin{equation}
\frac{v_{\rm j}\Gamma}{r^2}\frac{\partial (r^2 N_{i,{\rm c}})}{\partial r}-\frac{\partial\left(b_iN_{i,\rm c}\right)}{\partial E}+\frac{N_{i,{\rm c}}}{T_{i,\rm d}}=Q_{i,{\rm c}},\label{transportconvection}
\end{equation}   
{where the convection term is the first one on the left member and $r$ is the radius in spherical coordinates with the origin in the BH.} The term of decay is omitted for protons and it is kept in the case of pions and muons in the form $T_{i,\rm d}(E_i)= T_{i,\rm d}^0\left(\frac{E_i}{m_ic^2}\right)$, where $T_{i,\rm d}^0$ is the {lifetime} of the particle at rest. We solve Eq. (\ref{transportconvection}) applying the method of the characteristic curve as described in the appendix \ref{app.conesolution}. 

\begin{figure}[htbp] \label{Npcone}
\begin{center}
\includegraphics[width=.5 \textwidth,trim=3 15 20 0,clip]{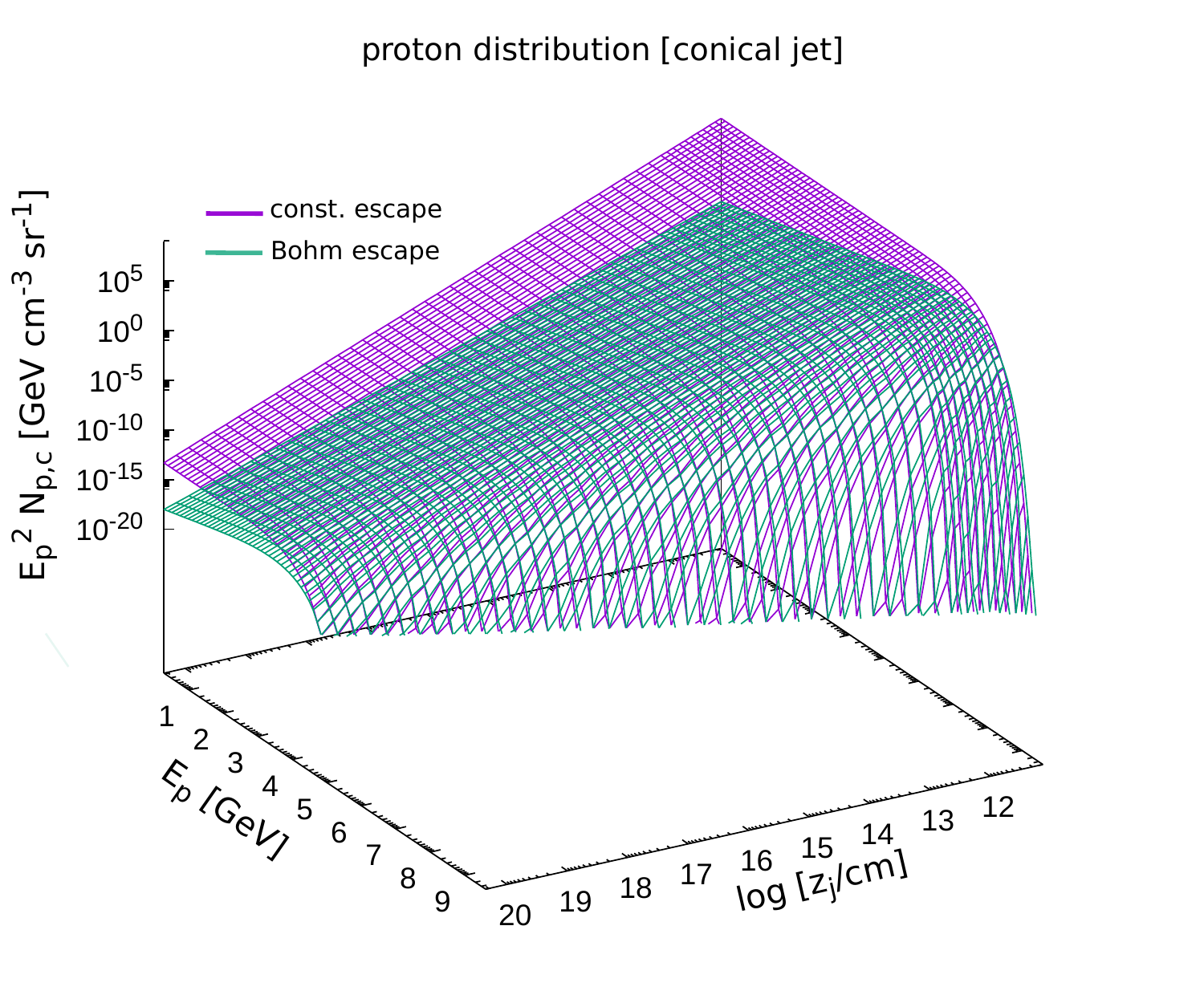}
\end{center}
\caption{Proton distributions at the conical part of the inner jet \textbf{for a constant escape rate from the base zone (magenta) and for a Bohm escape rate (green)}. The results correspond to a MQ at redshift $z = 8$ and  age $t_{\rm MQ} \sim 6.7\times 10^4 {\rm yr}$. \label{fig:Npcone}}
\end{figure}

In the case of protons, $T_{p,\rm d}\rightarrow \infty$, and after integrating and simplifying Eq.(\ref{transporEQconv_sol}), we obtain
\begin{multline}
N_{p,{\rm c}}(r,E_p)=\frac{\Delta V_{\rm b}N_{p,{\rm b}}(E'(r_{\rm eoi}))t^{-1}_{\rm esc}}{\pi R^2_{\rm j} v_{\rm j}}  H\left(r_{\rm eoi}-r_{\rm min} \right)\\ \times \left(\frac{r_{\rm eoi}}{r}\right)^{4+C_a}
\left(\frac{(1+C_a)r^{2+C_a}}{(r^{1+C_a}-r_{\rm eoi}^{1+C_a})C_bE_p- (1+C_a)r\,r_{\rm eoi}^{1+C_a}}\right)^2
\end{multline}
This result is shown in Fig. \ref{fig:Npcone} as a function of the position and the energy in the case of a MQ of age $t_{\rm MQ}= 6.7 \times 10^4{\rm yr}$ at redshift $z=8$. {As it can be seen, the distribution in the case of a Bohm diffusion escape is lower and with a flatter dependence on the energy. This is because the Bohm escape is proportional to the energy and hence this dependence affects the injection at the conical region. }

The dominant neutrino production process for such protons is through $pp$ interactions. {This is because the electrons injected in the base zone radiate practically all their power there before they can be injected in the conical zone. In other words, the escape rate for electrons at the base zone is orders of magnitude below the synchrotron cooling rate (see Fig. \ref{fig:pecool}, top-right panel), meaning that the electrons that can escape to be injected in the conical region carry only a negligible power. Therefore, there is no significant electron synchrotron radiation to act as target for $p\gamma$ interactions in the conical region.} 

The corresponding distributions of the produced pions and of the muons generated by pion decays are found using Eq. (\ref{transporEQconv_sol}) along with the injection given by
\begin{equation}
Q_{i,{\rm c}}(r,E)=  Q_i(z_{\rm j},E)+ \frac{\Delta V_{\rm b}}{\Gamma\pi R_{\rm j}^{2}(z_{\rm eoi})} N_{i,\rm b}(E) t_{p,\rm esc}^{-1} \delta(z_{\rm j} - z_{\rm eoi}), 
\end{equation}
where { $``i"$ refers to pions ($i=\pi$) and muons ($i=\mu$)}. The first term accounts for the injection produced along the jet and the second term corresponds to the injection due to the escape from the base zone. The obtained distributions of pions and muons are to be used in the calculation of the neutrino emission, as discussed below.

\subsection{Distribution of relativistic particles at the external zone}

Since we are interested in capturing all the relevant neutrino producing processes that can be triggered by high energy protons accelerated in Pop III MQs, and, in particular, taking into account that escape is dominant in the outermost zone of the system, we simply consider the injection of the protons escaping from the shell into an external zone at the IGM. In this way, we  account for the possibility that protons that are accelerated at the bow shock and escape from the shell could, in turn, generate neutrinos by $p\gamma$ interactions on the CMB if they are energetic enough to produce pions.

The corresponding {energy} density of CMB photons is \cite{ruffini2016}:
\begin{eqnarray}
n_{\rm ph,CMB}(z,E_{\rm ph})=\frac{8\pi E_{\rm ph}^2}{(hc)^3\left[{\rm exp}\left({\frac{E_{\rm ph}}{k_{\rm B}T_0(1+z) }}\right)-1\right]} \ ,
\end{eqnarray}
where $T_0=2.725\,{\rm K}$.
  Taking the size of this external zone to be $\Delta z_{\rm ext} = 10 \ \rm{Mpc}$ is adequate to consider the CMB as constant and homogeneous within. And this also leads to an escape rate which is lower than the $p\gamma$ cooling rate for energies above $\sim 5\times 10^{9}{\rm GeV}$, where pion production is activated. This can be seen in the left panel of Fig. \ref{fig:pcoolext} for MQs of age $t_{\rm MQ}=2\times 10^{4}$yr at a redshift $z=8$.

In order to obtain the proton distribution of this zone, we solve Eq. (\ref{TransportEQ}) with $t_{\rm cool}^{-1} = t_{p\gamma,\rm CMB}^{-1}$, $t_{\rm esc}^{-1} = c/\Delta z_{\rm ext}$, and:
\begin{equation} \label{qpext}
Q_{p,\rm ext}(E) = \frac{\Delta V_{\rm bs} }{\Delta V_{\rm ext}} N_{p,\rm bs}(E) t_{\rm esc, \rm bs}^{-1},
\end{equation} 
{where $\Delta V_{\rm ext}$ is the volume of the external zone. Eq. (\ref{qpext})} ensures that the total power injected matches the total power that escapes from the shell carried by protons. {The resulting distributions of protons are shown in the right panel of Fig.\ref{fig:pcoolext} for various redshifts, $t_{\rm MQ}\simeq 6.7\times 10^4{\rm yr}$. {We also show the cases for a constant escape from the shell and for a escape term assuming Bohm diffusion. In the latter case, it can be seen that the proton distributions are below the ones corresponding to a constant escape except at the highest energies.}

The distributions of secondary pions and muons are obtained following the procedure described above and are used to compute the expected neutrino output, as is discussed in the next section.

\begin{figure*}[htbp] \label{pcoolext}
\begin{center}
\includegraphics[width=0.95 \textwidth,trim=26 550 12 0,clip]{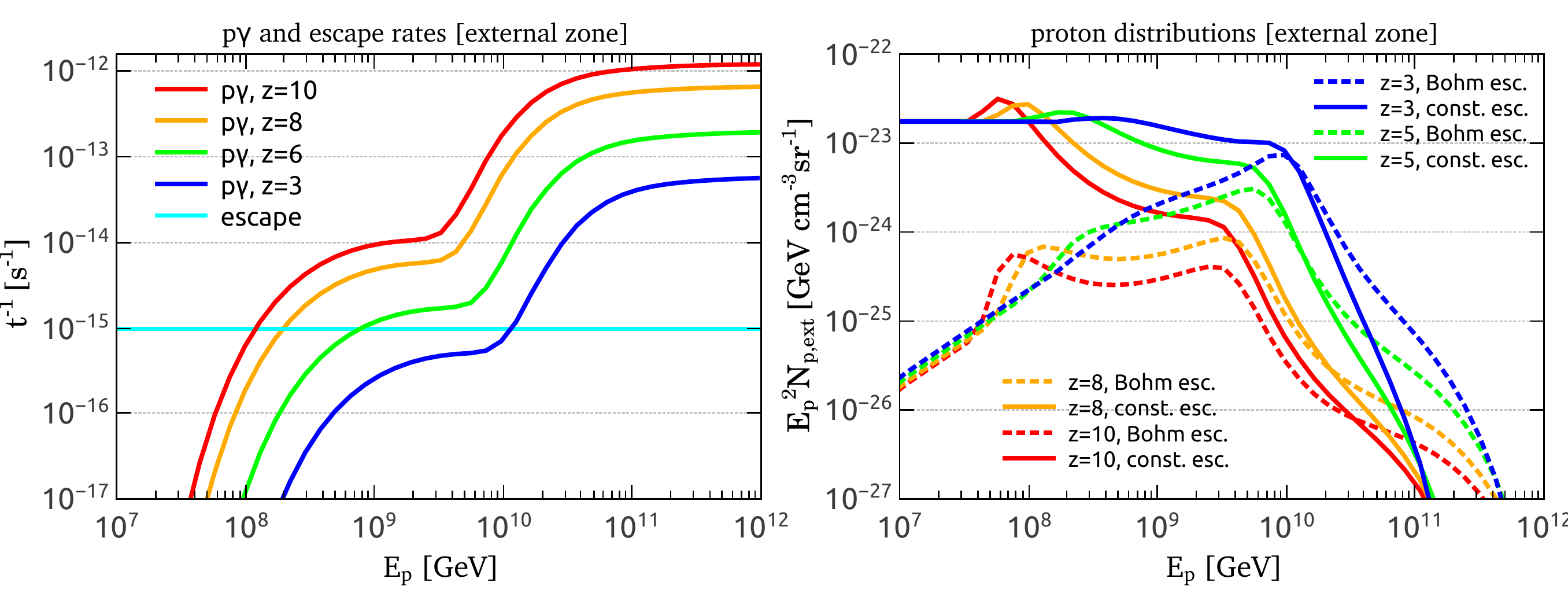}
\end{center}
\caption{{Proton cooling rates at the external zone for Pop III MQs of age $t_{\rm MQ}\simeq 6.7\times 10^{4}{\rm yr}$ at different redshifts (left panel), and the corresponding proton distributions (right panel)} {for a constant escape rate from the shell (solid lines) and for a Bohm escape rate (dashed lines)}. \label{fig:pcoolext}}
\end{figure*}

\section{Neutrino and electromagnetic emission \label{DNFsec}}

In each zone considered in the model, $pp$ and $p\gamma$ interactions lead to the production of charged pions, which decay to neutrinos and muons, and the latter also decay yielding neutrinos. The accompanying broadband photon emission co-produced by the high energy particles in our model is also computed consistently to check that it is not in conflict with any existing bound, as we show below.

The emissivity $\nu_\mu+\bar{\nu}_\mu$ from direct pion decays can be obtained following Ref. \citep{lipari2007}:
\begin{equation}
Q_{\pi\rightarrow\nu_\mu}(E_\nu)= \int_{E}^{\infty}dE_\pi
T^{-1}_{\pi,\rm d}(E_\pi)N_\pi(E_\pi)
\frac{\Theta(1-r_\pi-x)}{E_\pi(1-r_\pi)},
 \end{equation}
where {$\Theta(x)$ is the step function}, $x=E_\nu/E_\pi$, and the {pion lifetime} is $T_{\pi,\rm d}=2.6 \frac{E_\pi}{m_\pi c^2}\times 10^{-8}{\rm s}$.
The contribution from muon decays to $\nu_\mu+\bar{\nu}_\mu$ is {\citep{lipari2007}}
\begin{multline}
Q_{\mu\rightarrow\nu_\mu}(E_\nu)= \sum_{i=1}^4\int_{E}^{\infty}\frac{dE_\mu}{E_\mu} T^{-1}_{\mu,\rm
d}(E_\mu)N_{\mu_i}(E_\mu)  \\ \times \left[\frac{5}{3}-
3x^2+\frac{4}{3}x^3 +\left(3x^2-\frac{1}{3}-\frac{8x^3}{3}\right)h_{i}
\right],
\end{multline}
where $x=E_\nu/E_\mu$, the muon lifetime is $T_{\mu,\rm d}=2.2\frac{E_\mu}{m_\mu c^2}\times 10^{-6}{\rm s}$, $\mu_{1,2}=\mu^{-,+}_L$ and
$\mu_{3,4}=\mu^{-,+}_R$. {Here, L and R indicate the helicity} of the muons, that is $h_i=1$ for right-handed and $h_i=-1$ for left-handed muons. 
As for $\nu_e+\bar{\nu}_e$ , the emissivity from the decay of muons is given by {\citep{lipari2007}}:
\begin{multline}
Q_{\mu\rightarrow\nu_e}(E_\nu)= \sum_{i=1}^4\int_{E_\nu}^{\infty}\frac{dE_\mu}{E_\mu} T^{-1}_{\mu,\rm
d}(E_\mu)N_{\mu_i}(E_\mu,t) \\ \times \left[2-
6x^2+4x^3 +\left(2- 12x+ 18x^2-8x^3\right)h_{i}\right].
\end{multline}

The higher the Lorentz factors of the plasma in the different emission zones, the more boosted in the direction of the jet the observed flux would be. Certainly, counter-jets are de-boosted. Nevertheless, we still account for their contributions, since they can be significant, particularly in the cases of the shell and external zone, where bulk velocities are lower. 

The comoving emissivities above can then be transformed to the local frame at rest with the central BH to give 
\begin{equation}
Q_{\nu}' (E_{\nu}')= D_{i_{\rm j}} \ Q_{\nu} \left(\frac{E'_{\nu}}{D_{i_{\rm j}}}\right)+D_{\pi- i_{\rm j}} \ Q_{\nu} \left(\frac{E'_{\nu}}{D_{\pi-i_{\rm j}}}\right),\label{QnuBH}
\end{equation}
where $E_{\nu}' = D_{i_{\rm j}} E_{\nu}^{\rm com}$ is the energy in the BH frame and the Doppler factor corresponding to a viewing angle $i_{\rm j}$ is
\begin{equation} \label{dopplerfact}
D_{i_{\rm j}} = \left[\Gamma_j (1-\beta_j \cos i_{\rm j} )\right]^{-1},
\end{equation}
{and $\beta_j$ is the bulk velocity of the zone $j$ in units of $c$.}
 {We remark that the first term on the right member of Eq. (\ref{QnuBH}) includes the contribution of the jet, whereas the second term accounts for the counter-jet contribution.}

The neutrino spectrum, corresponding to one MQ at redshift $z$ for which the angle of the jet with the line of sight is $i_{\rm j}$ can be computed as:
\begin{equation}
\frac{dN_{\nu}'}{dE_{\nu}' d\Omega'} = Q_{\nu}'(E_{\nu}') \ dV \ d t_{\rm MQ}
\end{equation}

{In order to compute the diffuse neutrino flux due to all possible Pop III MQs that existed along the history of the universe, we consider the rate of their formation per unit mass to be a fraction of the corresponding rate of Pop III star formation:}
 \begin{equation} \label{RMQ}
 \frac{dR_{\rm MQ}(z)}{dM}=f_{\rm BH} \ f_{\rm bin}  \frac{dR_{\rm PopIII}(z)}{dM},
 \end{equation}
 {which represents the number of MQs generated per unit time, per unit volume and per unit mass of the Pop III stars produced ($M$). We consider that a fraction $f_{\rm BH}\simeq 0.9$ of stars with masss greater than $M_{\rm min,MQ}=50M_\odot$ produced BHs of about half of its mass, according to Ref.\citep{heger2002}, and a fraction $f_{\rm bin}\simeq 0.5$ were part of a close binary system \citep{stacy2013}.}
{At this point it is important to notice that the distribution with the mass of Pop III stars is still unknown, although there is certain concensous that it might be top-heavy according to recent simulations \citep{liu2020}, i.e., the total mass generated in stars is dominated by the contributed by the most massive ones. This corresponds to a distribution $\frac{dR_{\rm PopIII}(z)}{dM}\propto M^{-b}$, with $b=(0-2)$, being the most optimistic case the one of a flat distribution ($b=0$) \citep{hirano2016}, which leads to a greater number of high mass systems and hence to a higher neutrino emissivity overall.  }

  {We proceed to normalize, at each redshift $z$, the distribution $ \frac{dR_{\rm PopIII}(z)}{dz}$ using the total mass generated in Pop III stars $\dot{M}_{\rm PopIII}(z)$ according to Ref. \cite{desouza2011},
\begin{equation}
\int_{M_{\rm min}}^{M_{\rm max}}M\frac{dR_{\rm PopIII}(z)}{dM}dM= \dot{M}_{\rm PopIII}(z),
\end{equation}  
where we suppose that the possible range of masses for the stars is $M_{\rm min}\simeq 0.1 M_{\odot}$ and $M_{\rm max}= 100\,M_\odot$.  The total generation rate of Pop III MQs created by the evolution of stars with masses above $M_{\rm min,MQ}$ is a fraction of the total generation rate of Pop III stars $\dot{M}_{\rm PopIII}(z)$.} The latter is shown for illustration in Fig.\ref{fig:RMQ} and we also show it weighted by $H_0|\frac{dt}{dz}|$, where 

\begin{equation} \label{dtdz}
 \left|\frac{dt}{dz}\right|= \frac{1}{H_0(1+z)\sqrt{(1+z)^3\Omega_m+ \Omega_\Lambda}},
\end{equation}
with $H_0=70\, {\rm km\, s^{-1}Mpc^{-1}}$, $\Omega_m=0.315$, $\Omega_\Lambda= 0.685$.

\begin{figure}[htbp] 
\begin{center}
\includegraphics[width=.49 \textwidth,trim=0 12 0 0,clip]{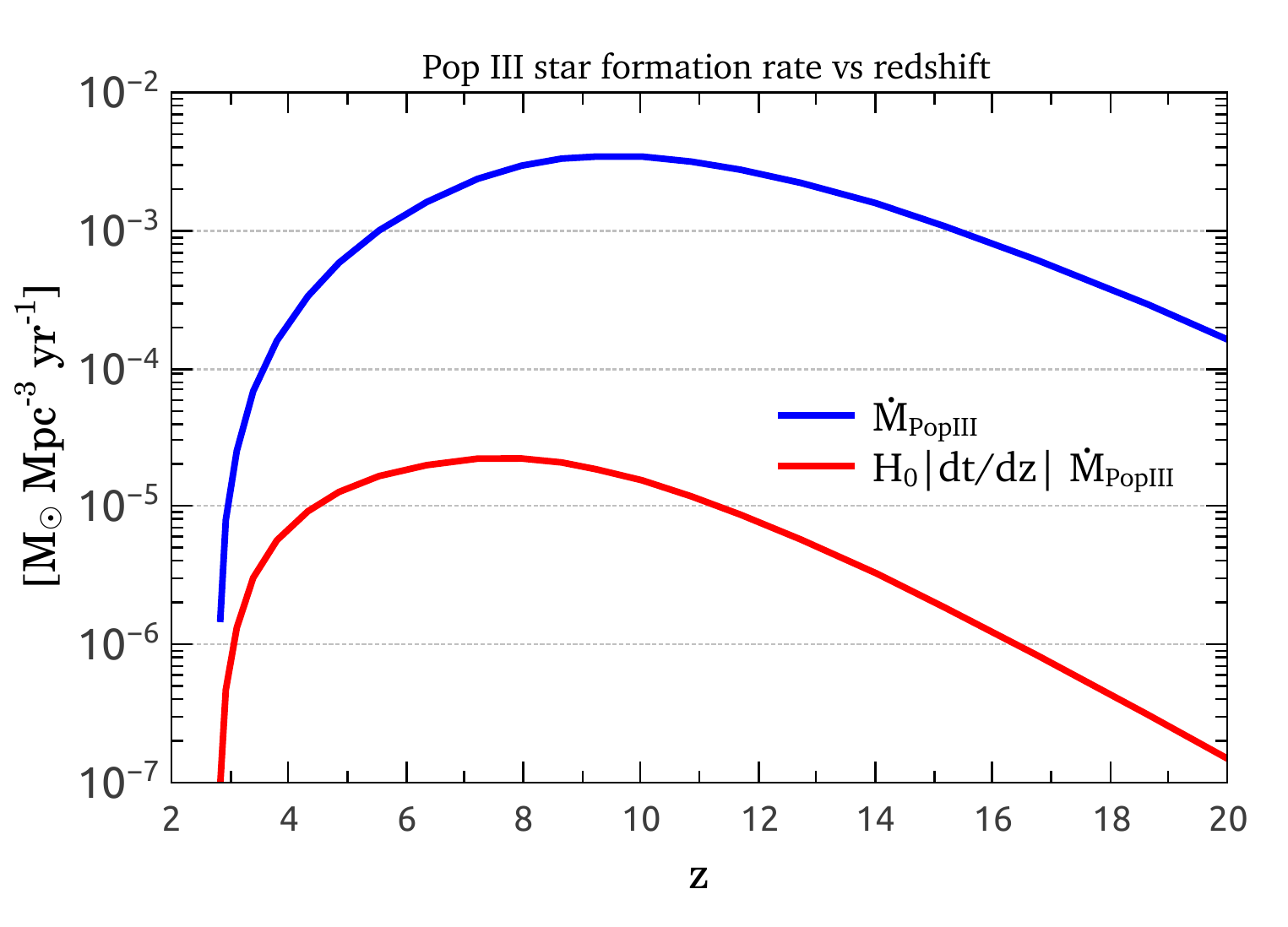}
\end{center}
\caption{{Formation rate of Pop III stars, $\dot{M}_{\rm PopIII}(z)$, in blue, adopted from Ref.\cite{desouza2011}. The product $H_0|\frac{dt}{dz}|\dot{M}_{\rm PopIII}(z)$ is shown in red, as this is useful for the calculation of the diffuse neutrino flux. \label{fig:RMQ}}}
\end{figure}

The differential density of the produced neutrinos using a similar expression to the given by Refs. \citep{andosato2004,murase2007}, i.e.,
\begin{equation}
\frac{dn_{\nu}(E_{\nu})}{dM} = \int d\Omega' \frac{dR_{\rm MQ}(z)}{dM}(1+z)^3 \left|\frac{dt}{dz}\right| dz \frac{dN_{\nu}'}{dE_{\nu}'d\Omega'} dE_{\nu}'(1+z)^{-3}. \label{Eqnnu}
\end{equation}

Eq. (\ref{Eqnnu}) accounts for the contributions of the jet and counter-jet, assuming that the orientation is distributed isotropically, that is, $\frac{dR_{\rm MQ}}{dM}$ is independent of $i_{\rm j}$. {According to Fig. \ref{fig:RMQ}, it can be concluded that the dominant neutrino contribution arises for reshifts $z\approx 7-8$, where $H_0|\frac{dt}{dz}|\dot{M}_{\rm PopIII}(z)$ peaks. }

Integration over the total MQ life $T_{\rm MQ}$, the solid angle, and volume of the emitting zone $j$, yields, in units of $\rm [\rm energy^{-1}]$, {the spectrum of the muonic flavor of neutrinos and antineutrinos, $\nu_\mu + \bar{\nu}_\mu$ :
}
\begin{multline}
\frac{dN_{\nu}^{'}}{dE_{\nu}^{'}} = 4 \pi \int_{0}^{\Delta V_{j}} dV_j \int_{0}^{T_{\rm MQ}} dt_{\rm MQ} \int_{0}^{\frac{\pi}{2}} di_{j} \sin(i_{j}) \times \\ 
   \left[ Q'_{\nu_\mu}(E_\nu') \, P_{\nu_\mu\rightarrow \nu_\mu} +
Q'_{\nu_e}(E_\nu') \, P_{\nu_e\rightarrow \nu_\mu}\right]  . \label{nuspectrum} 
\end{multline}
{Here, $P_{\nu_{\mu}\rightarrow \nu_{\mu}}\simeq 0.453$ is the probability that the generated $\nu_\mu$ or $\bar{\nu}_\mu$ keep the same flavor, and $P_{\nu_{e}\rightarrow \nu_{\mu}}\simeq 0.171$ is the probability that $\nu_e$ or $\bar{\nu}_e$ oscillate into $\nu_\mu$ or $\bar{\nu}_\mu$.  These probabities are derived from the unitary mixing matrix $U_{\alpha j}$, which is determined by three mixing angles, $\theta_{12}\simeq 33.4^\circ$, $\theta_{13}\simeq 8.57^\circ$, and $\theta_{23}\simeq 49^\circ$, and the CP-violating phase $\delta_{\rm CP}\approx 197^\circ$ \citep{esteban2020}. The values used for the probabilities correspond to a normal mass ordering of the massive neutrinos $(\nu_1,\nu_2,\nu_3)$, i.e., $m_1<m_2<m_3$.}

Considering that the emission from any redshift $z$ is the same in all directions, the differential neutrino density can be related to the differential neutrino flux with as $\frac{d\Phi_{\nu}}{dE_{\nu}} = \frac{c}{4\pi} n_{\nu}$. Hence, the final expression for the diffuse neutrino flux originated in Pop III MQs is given by
\begin{equation}
\frac{d\Phi_\nu(E_\nu)}{dE_\nu}= \frac{c}{4\pi}\int_{z_{\rm min}}^{z_{\rm max}}{dz \left|\frac{dt}{dz}\right| \int_{M_{\rm min,MQ}}^{M_{\rm max,MQ}} dM \frac{dR_{\rm MQ}(z)}{dM}\frac{dN_\nu'}{dE_\nu'}},\label{DNF_eq}
\end{equation}
{where $M_{\rm min,MQ}=50\,M_\odot$ and $M_{\rm max,MQ}=100\,M_\odot$ limit the range of masses of Pop III stars supposed to lead to the formation a MQs, and  $z_{\rm  min}= 3$ and $z_{\rm max}=25$ indicate the limiting values of redshift along which Pop III MQs were distributed according to $\dot{M}_{\rm Pop III}(z)$.} 

\begin{figure*}[htbp] 
\begin{center}
\includegraphics[width=\textwidth,trim=2 0 13 0,clip]{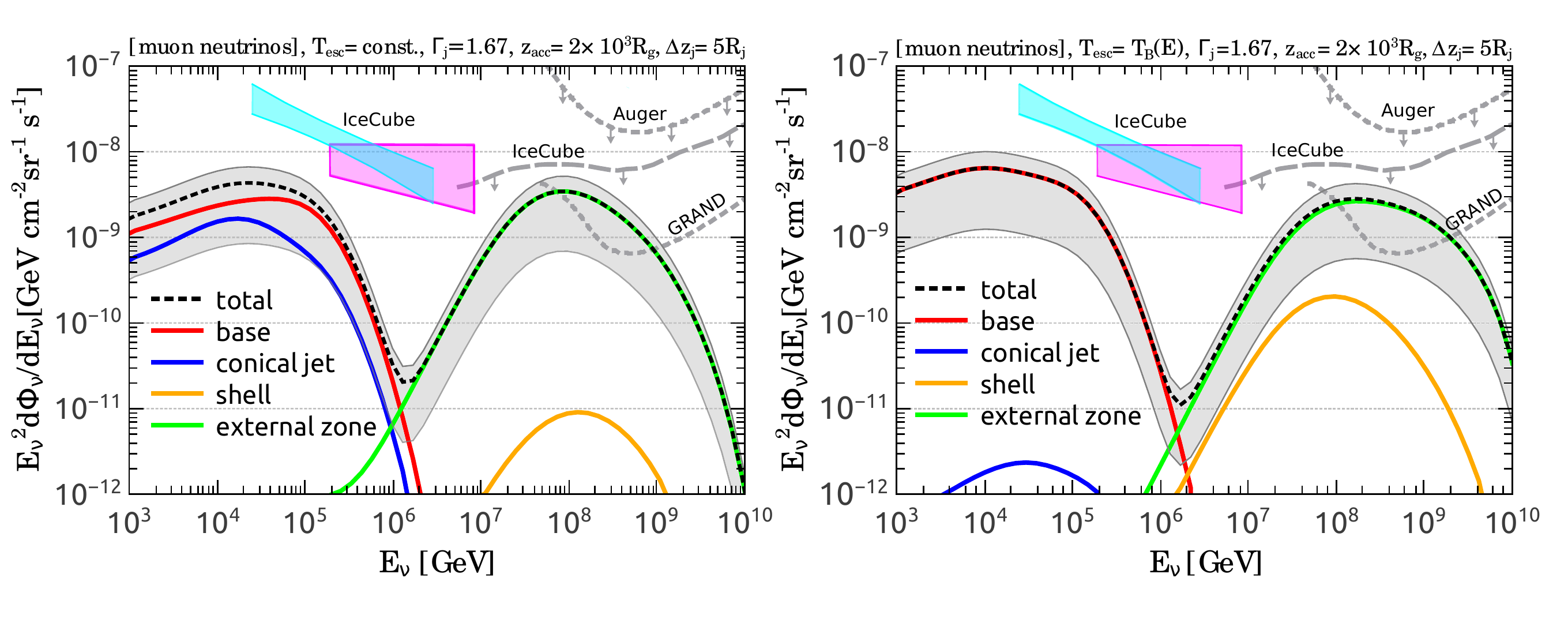}
\end{center}
\caption{Diffuse neutrino flux including the major contributions {in the cases of a constant escape rate (left panel) and a Bohm escape rate (right panel). The shaded region corresponds to varying the index $b$ characterizing the mass distribution of MQs between $b=0$ (highest flux) and $b=2$ (lowest flux).} \label{fig:DNF}}
\end{figure*}

\begin{figure*}[htbp]
\begin{center}
\includegraphics[width= \textwidth,trim=1 9 0 0,clip]{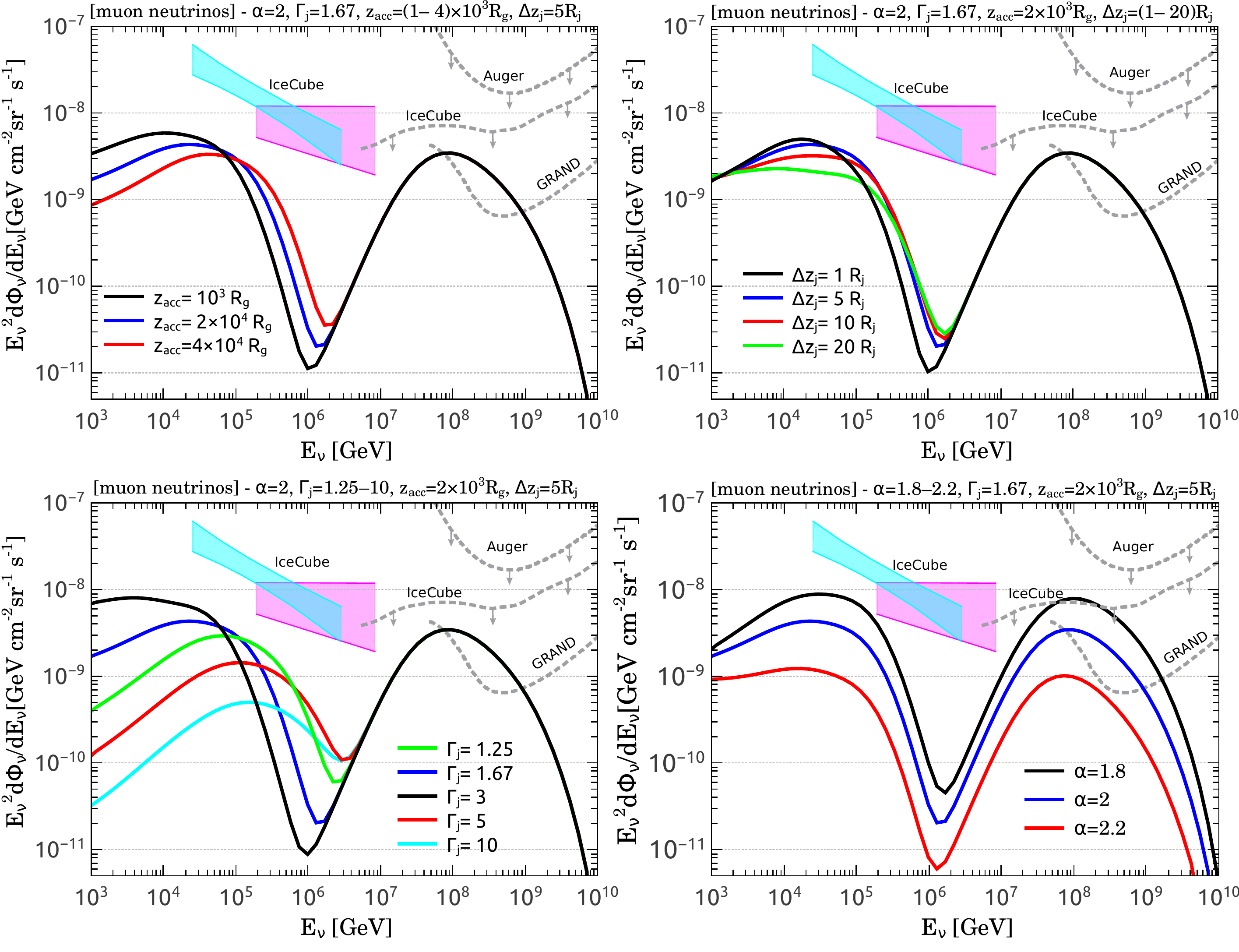}
\end{center}
\caption{Diffuse neutrino flux resulting with different combinations of parameters in the case of a constant escape rate. \label{fig:DNF_swp}}
\end{figure*}

Taking into account the dominating cooling processes described above, we can make some simple {order-of-magnitude} estimates of the neutrino flux that could be expected at the Earth. 
As it can be seen from Fig. \ref{fig:pecool}, $p\gamma$ interactions at the base zone are most effective at high energies, $E_p\sim 10^7$ GeV, and escape dominates otherwise for the adopted parameters {in the case of a constant escape rate}. The escaping protons cool dominantly by adiabatic expansion, but they can still undergo $pp$ interactions along the rest of the conical part of the jet. Considering that, on average, $\sim 20\%$ of the energy of the parent proton goes to the produced neutrinos \cite{atoyandermer2003}, we can estimate that the average power carried by the final neutrinos generated in the inner jet to be $L_{\nu,\rm jet} \approx 0.01 L_{p,\rm b}$ {for a neutrino energy range between $E'_{\nu,1}\approx 1.5\times 10^3\,{\rm GeV}$  and $E'_{\nu,2}\approx 1.5\times 10^6\,{\rm GeV}$}. This estimation accounts for the fact that $p\gamma$ interactions do not dominate over the whole mentioned energy range, but only for the most energetic protons.} Considering a typical lifetime $T_{\rm MQ}\approx 2\times 10^5{\,}{\rm yr}$ for Pop III MQs, a simplistic $\sim {E'}_\nu^{-2}$ {single-source} spectrum of neutrinos of all flavors can be obtained as
\begin{eqnarray}
 \left.\frac{dN'_\nu}{dE'_\nu}\right|_{\rm jet}&\approx& {L_{\nu,\rm jet} T_{\rm MQ}}\log\left(\frac{E'_{\nu,2}}{E'_{\nu,1}}\right){E'}_{\nu}^{-2} \\  &\simeq & 2.8\times 10^{52}{\rm GeV^{-1}}\left(\frac{L_{p,\rm b}}{L_i}\right)\left(\frac{T_{\rm MQ}}{T_{\rm 0.2\,Myr}}\right)\left(\frac{E'_\nu}{\rm GeV}\right)^{-2},\nonumber
\end{eqnarray}
where $L_i=5\times 10^{39}{\rm erg\,s^{-1}}$ and $T_{\rm 0.2\,Myr}=2\times 10^5{\rm yr}$.
Similarly, since the protons accelerated at the shell escape to the external zone, and those with energies $E_p\gtrsim 5\times 10^9{\rm GeV}$ photo-produce pions efficiently by interactions with the CMB, the power carried by the neutrinos generated is roughly $$L_{\nu,\rm ext} \approx 0.2\times L_{p,\rm bs}(E_p>5\times 10^9{\rm GeV})\simeq 0.02\times 5\times 10^{38}{\rm erg\, s^{-1}}\left(\frac{L_{p,\rm bs}}{0.1L_i}\right).$$ 
Therefore, an estimate for a typical spectrum of the neutrino produced in the external zone is
\begin{eqnarray}
 \left.\frac{dN'_\nu}{dE'_\nu}\right|_{\rm ext}&\approx & {L_{\nu,\rm ext} T_{\rm MQ}}\log\left(\frac{E'_{\nu,2}}{E'_{\nu,1}}\right){E'}_{\nu}^{-2} \\ &\simeq & 3.4\times 10^{52}{\rm GeV^{-1}}\left(\frac{L_{p,\rm ext}}{0.1\,L_i}\right)\left(\frac{T_{\rm MQ}}{T_{\rm 0.2\,Myr}}\right)\left(\frac{E'_\nu}{\rm GeV}\right)^{-2},\nonumber
\end{eqnarray}
with $E_{\nu,1}=2.5\times 10^{8}{\rm GeV}$ and $E_{\nu,2}= 2.5\times 10^9{\rm GeV}$.

{The integrals on $M$ and $z$ of Eq.(\ref{DNF_eq}) can be estimated making the rough approximation that the rate of generated MQs is such that 
$$H_0\left| \frac{dt}{dz}\right|\frac{dR}{dM}\Delta M \sim f_{\rm BH}f_{\rm bin}\frac{\dot{M}_{\rm PopIII}}{50M_{\odot}}\simeq 1.5\times 10^{-7} {\rm Mpc^{-3}yr^{-1}}$$}
for redshifts between $z_1=5$ and $z_2=10$ (see Fig. \ref{fig:RMQ}). Assuming that flavor mixing leads to an approximate equal ratio for the three neutrino flavors,  the contributions to the diffuse flux from the inner jet and the external zone are roughly given by

\begin{multline}
\left.E_\nu^{2}\frac{d\Phi_\nu}{dE_\nu}\right|_{\rm jet}\approx  8\times 10^{-9}{\rm GeV \, cm^{-2}sr^{-1}s^{-1}}\left(\frac{L_{p,\rm b}}{L_i}\right)\left(\frac{T_{\rm MQ}}{T_{\rm 0.2\,Myr}}\right)  \\ {\rm for}  \   250{\rm GeV}\lesssim  E_\nu\lesssim 2.5\times 10^5 {\rm GeV}  \nonumber   
\end{multline}
\begin{multline}
\left.E_\nu^{2}\frac{d\Phi_\nu}{dE_\nu}\right|_{\rm ext} \approx  10^{-8}{\rm GeV \, cm^{-2}sr^{-1}s^{-1}}\left(\frac{L_{p,\rm bs}}{0.1\,L_i}\right)\left(\frac{T_{\rm MQ}}{T_{\rm 0.2\,Myr}}\right)  \\ {\rm for}  \   4\times 10^{7}{\rm GeV}\lesssim  E_\nu\lesssim 4\times 10^8 {\rm GeV} \nonumber.
\end{multline}


While useful as order of magnitude estimations, these expressions clearly do not account for the exact dependence of the particle distributions, injections, and intervening cooling rates, so that, for instance, the effect of synchrotron losses by pions and muons at the inner jet were not included at that point. 
{Another important issue is accounting for MQs with different BH masses which arise, as explained above, by the gravitational collapse of a Pop III stars with masses between $50M_\odot$ and $100M_\odot$ in binary systems. We address this point by considering that the power injected in relativistic particles is proportional to the BH mass, and hence to $M$. Therefore neutrino emission is also proportional to $M$ for all the processes except for $p\gamma$ at the jet base, since there the target photons correspond to synchrotron emision by the electrons, which is also proportional to $M$. Hence, neutrino production the jet base is considered to scale as $\propto M^2$. We apply these scalings to perform the integration over $M$ in Eq.(\ref{DNF_eq}) making use of our central result obtained for $M_{\rm BH}=30M_\odot$, i.e. for a Pop III star mass $M\simeq 60\,M_\odot$. This can be performed for different cases of mass distributions $\frac{dR_{\rm MQ}}{dM}\propto M^{-b}$, with $b=(0-2)$ as discussed above. In Fig. \ref{fig:DNF} we show the results obtained with the full numeric code for the diffuse neutrino flux of $\nu_\mu+\bar{\nu}_\mu$ in the case of $b=1$, and the gray shaded region indicates the possible range of the flux values between the lowest flux corresponding to $b=2$ and the highest one for $b=0$.  
We also show individually the most significant contributions among the different emission zones considered for the cases of constant escape rates (left panel) and for Bohm escape rates (right panel).}  
  We also include the fit obtained for IceCube data \citep{aartsen2015,aartsen2017}, as well as the upper limits of higher energy neutrinos given by {Auger} \citep{pierreauger2018} and IceCube \citep{aartsen2018}. {For reference, we also show the expected sensitivity for GRAND \citep{GRAND2018}, but other planed detectors will be sensible to UHE neutrinos as well, such as IceCube-Gen2 \citep{icecubegen2}, PUEO \citep{pueo}, RNO-G \citep{rno2019}, Trinity \citep{trinity2019}, and BEACON \citep{beacon2020}.} 
 
 In Fig. \ref{fig:DNF_swp}, we plot the results corresponding to the diffuse neutrino flux if four key parameters of the model are varied, {adopting for illustration a constant escape rate}. In the top left panel, we show the fluxes obtained for different values of the position of the emitter in the jet base ($z_{\rm acc}$), while in the top right panel, the size of the base zone ($\Delta z_{\rm b}$) is varied. Likewise, the resulting flux is shown for different values of the jet Lorentz factor ($\Gamma$) in the bottom left panel, and with different values of the index of injection of primary particles ($\alpha$). Although for simplicity we have kept the ratio of magnetic to kinetic energy at the base as constant ($q_{m}=5\times 10^{-3}$), varying $z_{\rm acc}$ leads to different values of the magnetic field at the base zone, since its size is set in reference to the expanding jet radius. Then, in the top left panel of Fig.\ref{fig:DNF_swp}, $\Delta z_{\rm b}=5R_j\simeq 0.5 z_{\rm acc}$ is assumed in the three cases, and we obtain $B_{\rm acc}\simeq 2.1\,{\rm G}\frac{z_{\rm acc}}{R_g}$, with $R_g\simeq 4.4\times 10^8{\rm cm}$. In the top right panel, we fix $z_{\rm acc}=2\times 10^4 R_g$ and changing the size of the base zone basically modifies linearly the escape rate and the $p\gamma$ cooling rate. Therefore, for instance, for the highest value considered ($\Delta z_{\rm b}=20 R_g$), both rates are low leading to a less effective neutrino production  in comparison with the other cases for smaller sizes. In particular, for the smallest value adopted ($\Delta z_{\rm b}=R_g$), it can also be seen that the maximum neutrino energy is lower, and this is because the acceleration rate is the same for all the cases of that panel and the maximum proton energy is correspondingly lower for the high rates of escape and $p\gamma$ collisions. We note that, given the values of the magnetic field considered at the base in general, the electron cooling is so fast that no significant synchrotron emission takes place outside the injection region at the jet base. Therefore, if the volume of this zone is increased, the density of synchrotron decreases and $p\gamma$ become less effective.
 
 In the bottom left panel of Fig.\ref{fig:DNF_swp}, it can be seen that the contribution from the inner jets decreases as the bulk Lorentz factor of the jet increases. This can be understood as a consequence of the fact that under the assumptions made, the neutrino emissivity in the comoving frame is $Q_\nu\propto \Gamma^{-1}$, as is shown in Appendix \ref{app.pgamma}. Therefore, when transformed to the BH frame, Eq.(\ref{QnuBH}) implies that $Q_\nu'\propto \Gamma^{-2}$ and this is reflected in the final possible fluxes to arrive at the Earth. Varying $\Gamma$ also implies varying the magnetic field at the base, since the magnetic energy is proportional to the kinetic one, and the latter is $\propto \Gamma(\Gamma-1)$ (Eq.\ref{rhokin}). Therefore, for $\Gamma=1.25$, $B_{\rm acc}\approx 9.5\times 10^4{\rm G}$, and for $\Gamma=10$ we have $B_{\rm acc}\approx 4.4\times 10^{4}{\rm G}$.   
In the bottom right panel, we show the diffuse neutrino flux obtained for other values of the spectral index of primary particle injection: $\alpha=1.8$ and $\alpha=2.2$. As it can be seen, the prospects for detection fall for steeper injections.    

\begin{figure}[htbp] \label{flav}
\begin{center}
\includegraphics[width=.48 \textwidth,trim=0 0 0 0,clip]{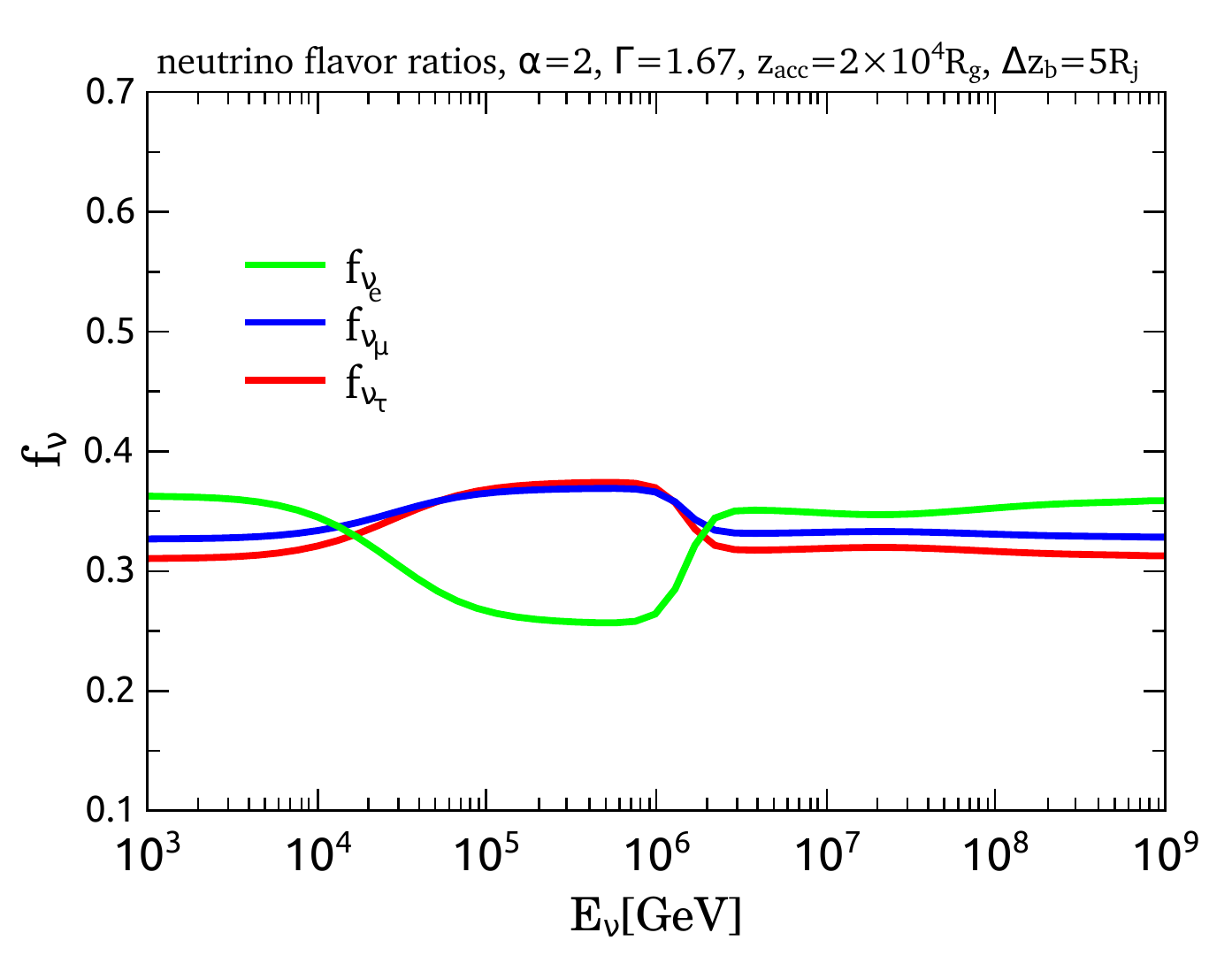}
\end{center}
\caption{Neutrino flavors ratios as a function of energy in the case of a Bohm escape rate. \label{fig:flav}}
\end{figure}

The flavor ratios of neutrinos has become an interesting observable which can bring information on the nature of the production mechanism operating at the sources \citep{beacom2005,mena2014,bustamante2015,icecubeflavor2015,bustamante2019}. In Fig. \ref{fig:flav}, we show the neutrino flavor ratios that are obtained within our model for Pop III MQs, {in the case of a Bohm escape rate, but the result is very similar for a the constant escape case}. The effect caused by the magnetic field is manifest for the energy window $\sim(3\times 10^4- 10^6){\rm GeV}$, where high energy muons at the inner jet are affected by synchrotron losses and a deficit of electron neutrinos are produced. For still higher energies, the flux is dominated by the contribution from the escaping protons interacting with the CMB, and no magnetic field effects are expected. This also happens for neutrino energies $E_\nu \lesssim 10^4{\rm GeV}$, for which synchrotron losses of pions and muons at the inner jet are not significant.

\begin{figure*}[htbp]
\begin{center}
\includegraphics[width=0.95\textwidth,trim=3 0 105 0,clip]{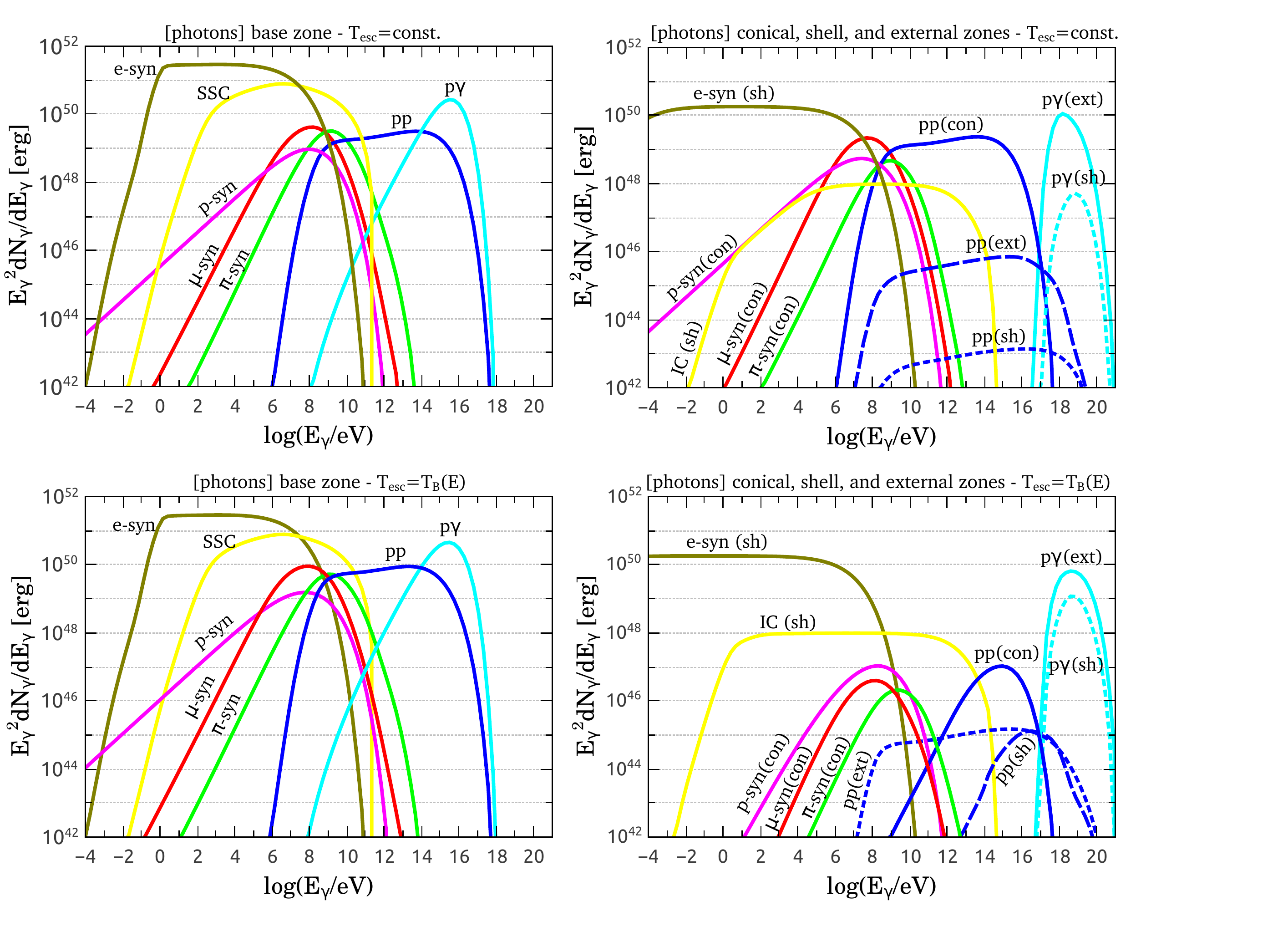}
\end{center}
\caption{Total photon spectra emitted by MQs at redshift $z=8$ in the case of a constant escape rate \label{fig:photonspectra}.}
\end{figure*}

\begin{figure}[htbp] 
\begin{center}
\includegraphics[width=.48 \textwidth,trim=0 0 0 0,clip]{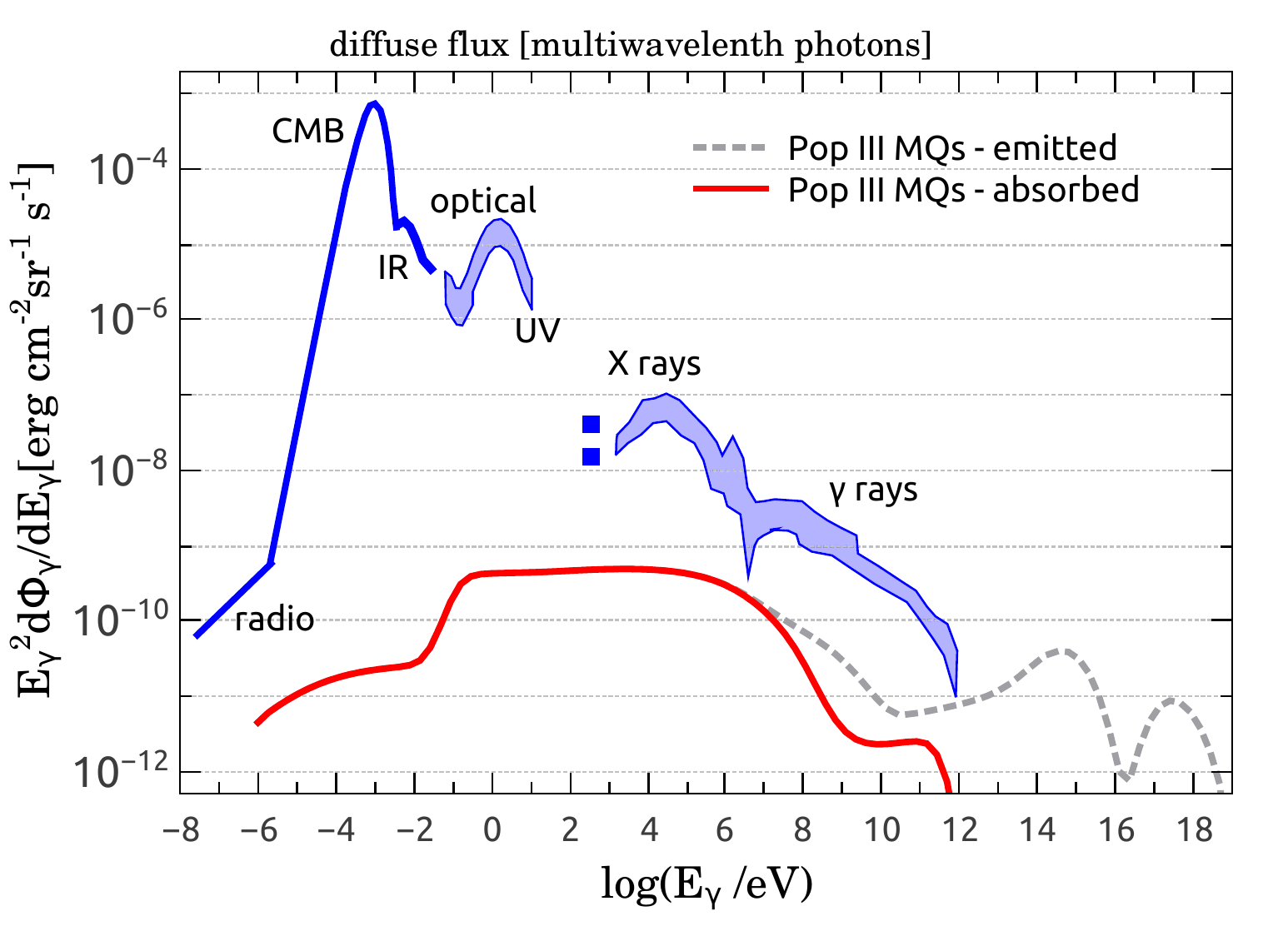}
\end{center}
\caption{Diffuse flux of multiwavelength photons from Pop III MQs in the case of a constant escape rate as compared to existing data of the  extragalactic photon background, adapted from Ref. \citep{cooray2016} \label{fig:DPF}.}
\end{figure}

For completeness, we compute the diffuse background flux of multiwavelength photons that are co-produced along with the neutrinos, using a expression analogous to Eq. (\ref{DNF_eq}). The main contributing processes are synchrotron emission, IC interactions, $pp$, and $p\gamma$ collisions, and we compute the corresponding emissivities following, e.g., Refs. \citep{vila2008,magnetic,lepto}. We show in Fig. \ref{fig:photonspectra} the photon spectra obtained for MQs at redshift $z=8$, where the corresponding to the base zone is presented on the left panel, and the most significant contributions from terminal jet are shown on the right panel. The diffuse flux obtained is shown in Fig. \ref{fig:DPF}, where it can be seen that the flux level is well below that of the extragalactic background of multiwavelength photons \cite{cooray2016}. We include the flux corrected by $\gamma\gamma$ absorption on the CMB and EBL through an exponential factor $e^{-\tau_{\gamma\gamma}}$, where the optical depth $\tau_{\gamma\gamma}$ is integrated following Ref. \cite{dominguez2010}.


\section{Discussion \label{discussion}} 

In this work, we have applied a model that allows to obtain a diffuse neutrino flux produced by a distribution of Pop III MQs during their lifetime at a wide range of redshifts ($ z = 3 - 25$). The flux of multiwavelengh photons is consistently computed and it is in agreement with observational data. 
As super-accreting sources and more massive than the typical galactic MQs, Pop III MQs should be capable of ejecting more powerful jets. We have adopted sets of parameters with values that are physically plausible for these systems, and at the same time favor particle acceleration and high energy neutrino production. 

In particular, we assumed a high acceleration efficiency and also that the injected power in electrons ($L_{e,j}$) is the same as that in protons ($L_{p,j}$) at each emission zone ``$j$" considered. We explored different combinations of parameters, varying the position and size of the emitter at base zone of the inner jet, the Lorentz factor of the jet, and the index of injected particles. Out of the possibilities mentioned, which involve plausible values for the parameters, we select the case shown in {right panel of} Fig. \ref{fig:DNF} as a representative one for efficient neutrino production. 
The main contributions to the diffuse neutrino flux arise at the inner jets for energies $\sim [10^4-10^6]{\rm GeV}$, and at the external zone for higher energies ($\sim [10^7 - 10^9] {\rm GeV}$). 

The inner jets are relevant sites for neutrino production for the following reasons: first, at the base of the jet the magnetic field derived is strong ($B\sim 10^4 {\rm G}$), which enhances the acceleration efficiency and also the production of the low energy photons generated by electron synchrotron, thus favoring $p\gamma$ interactions. And second, the target of cold protons at this region is highly dense, $n_p \sim 10^{12} \ {\rm cm}^{-3}$ at $z_{\rm j}=z_{\rm acc}$, which favors the $pp$ interactions at the jet base and at the conical part of the jet. The contribution from the inner jet still do not {reach the} level of the detected neutrino flux according to a global fit of IceCube \cite{aartsen2015},
and the best neutrino fit of astrophysical $\nu_\mu + \bar{\nu}_\mu $ \cite{aartsen2016}, at energies  $\sim 2\times 10^5  \rm GeV$. In order to have a higher neutrino flux, the typical jet power could be increased, and/or the typical MQ lifetime, but this would be inconsistent with the results obtained by simulations in Ref. \citep{inayoshi2017}. The mentioned parameters were taken from this reference an the fraction of power that is injected in relativistic particles was considered with the typical value $q_{\rm rel}=0.1$ as is commonly assumed in the similar models \citep{vila2008,bordas2009,sotomayor2019}, so trying to adopt still higher values for these parameters seems hard to justify.   

As for neutrino production at the terminal regions of the jet, we find that the dominant contribution arises from the shell, where due to a lower magnetic field and a larger size of the emission zone, the maximum proton energy can be as high as $\sim 10^{10}$GeV (see Fig.\ref{fig:pecool}), thus generating neutrinos peaking in the energy range $\sim(10^7-10^9)$GeV. However, since proton escape from the shell actually dominates over $p\gamma$ interactions within it, the great majority of the protons indeed escape and are injected into the IGM. The external zone considered allows to account for the possibility that further $p\gamma$ interactions with the CMB take place outside the MQs, and we found that these give the major contribution at the highest energy part of the obtained neutrino output.  	  
This contribution does not violate the upper limits given by Pierre Auger Observatory and IceCube, but could still be at the reach of  future detectors such as GRAND, as shown in Figs. \ref{fig:DNF} and \ref{fig:DNF_swp}. 
Furthermore, for energies from $\sim 10^8$ to $10^9 {\rm GeV}$, this contribution to the diffuse neutrino flux overlaps the energy range that the flux expected from cosmogenic neutrinos produced by the interaction of ultra-high energy cosmic rays (UHECRs) with photon targets from the CMB. Cosmogenic neutrinos are {sensitive} to the chemical composition of UHECRs, namely, their expected flux is higher for higher proton content in UHECR with respect to heavier nuclei. {On the other hand, since no significant heavy nuclei contribution is expected from Pop III MQs because these elements are released by supernova explosions and Pop III stars are the first generation of stars in the universe and have zero metallicity. This means that these stars basically burn hydrogen to helium so that heavier nuclei are not present in the accreeting matter, and hence can not be accelerated in jets of Pop III MQs.} {The simple approach applied to obtain this contribution is still adequate as long as over the interaction length, the photon background can be considered as constant, and this condition is satisfied. 
Since, as mentioned no significant contribution of heavy nuclei is present, hence it is in principle not necessary to account for a cascade of nuclear reactions. We also do not compute any electromagnetic cascade that would develop by interactions with the CMB. However, the emission by the dominant processes allows to conclude that there is no conflict with data (see Fig. \ref{fig:photonspectra}), and this is enough for the purposes of the present work.}

{Detailed studies for cosmogenic neutrino production account for in-source nuclear cascades \citep{biehl2017,morejon2019} to characterize the correct level of neutrino flux consistent with different chemical compositions. Therefore, if cosmic ray data finally established a chemical composition consistent with a very weak flux of accompanying cosmogenic neutrinos, and if future neutrino observations yield a diffuse signal above the predicted level, then the posibility that the sources of these neutrinos could be Pop III MQs should not be ruled out} . Conversely, in case of a future non-detection of the high energy part of the neutrino flux predicted, this would require that either Pop III MQs themselves did not generate at the rate here assumed, and/or that their efficiency for accelerating protons at their shells should be bound to a lower value than the assumed in this work. {For instance, lower values of the efficiency of acceleration $\eta$ would shift the bumps of the main contributions towards lower energies, and in the case of the external zone neutrino production by protons accelerated at the shell could even be supressed if the pion production threshold is not reached.}

Future neutrino observations with new generation instruments such as IceCube-gen2, {GRAND,  PUEO, RNO-G, Trinity, and BEAC} will be useful to probe the flux neutrinos from Pop III MQs at the highest energies. This will also help to obtain more accurate measurments of the flavor composition along an extense energy range, which would constrain neutrino producing models such as the presented in this work and yield more light on the origin of astrophysical neutrinos.

\appendix
\section{Solution of the inhomogeneous transport equation with convection and decay}\label{app.conesolution}
 Here we describe the steps followed to solve Eq.(\ref{transportconvection}) using the method of the characteristics. We assume the boundary condition $N_{\rm c} (r, E_i)\vert_{r \longrightarrow  0} = 0$, i.e., the escaping particles have a vanishing distribution at $r \ll z_{\rm eoi}$.
Since the dominant cooling processes are adiabatic expansion and synchrotron emission, we rewrite the transport equation as:
\begin{multline}
\frac{1}{r^2}\frac{\partial (r^2 N_{i,{\rm c}})}{\partial r}-  \left[C_a\frac{E}{r}+C_b\frac{E^2}{r^2}\right]\frac{\partial N_{i,{\rm c}}}{\partial E} \\ -
   \left[\frac{C_a}{r}+  \frac{2C_b E}{r^2} - \frac{C_c}{ E}\right]N_{i,{\rm c}}=\frac{Q_{i,{\rm c}}}{\Gamma v_{\rm j}},
\end{multline}
with the constants are given by:
\begin{eqnarray}
C_a &=& \frac{2}{3\Gamma}  \label{Cad}\\
C_b &=& \frac{4}{3}\left(\frac{m_e}{m_i}\right)^3 \sigma_T c \frac{B_0^2}{8\pi}\frac{z_{\rm acc}^2}{m_e c^2} \frac{1}{\Gamma v_{\rm j} m_p c^2}\label{Cbsyn} \\
C_c &=& \frac{m_i c^2}{T_0\Gamma v_{\rm j}}\label{Cddec}
\end{eqnarray}
The solution to the characteristic equation
\begin{equation} 
\frac{dE_i}{dr}= -C_a\frac{E_i}{r}- C_b\frac{E_i^2}{r^2},
\end{equation}
gives the characteristic curve
\begin{equation}
E'(r';r,E_i)= \frac{(1+C_a)E_i\, r'^{2}r^{1+C_a}}{(1+C_a)r'^{1+C_a}r^2+C_b E_i\,({r'^{1+C_a}}- r^{1+C_a})}.\label{EcharAdSyn}
\end{equation}
Using the curve corresponding to each pair of values $(r,E_i)$, we solve the following ordinary differential equation,
\begin{equation}
\frac{dN_{i,{\rm c}}}{dr'}=  \frac{Q}{\Gamma v_{\rm j}}+\left(\frac{C_a}{r'}+\frac{2C_bE'(r')}{{r'}^2} -\frac{C_c}{E'(r')}-\frac{2}{r'}\right){N_{i,{\rm c}}}.
\end{equation}
to obtain the particle distribution along the inner jet as:
\begin{multline}
N_{i,{\rm c}}(r,E_i)=\int_{r_{\rm ini}}^{r}dr'\frac{Q(r',E'(r'))}{\Gamma v_{\rm j}}  \times \\ \exp\left\lbrace\int_{r'}^r \frac{dr''}{E'(r'')r''^2}\left[{C_a r''E'(r'') }\right.\right. \\ \left.\left.{+ 2C_b\left(E'(r'')\right)^2-2E'(r'')r''-r''^2/T_{i,\rm d}}\right] \right\rbrace. 
\label{transporEQconv_sol}
\end{multline} 

Here, $r_{\rm ini}={\rm max}\left(r_{\rm acc},r_{\rm min}\right)$, where $r_{\rm min}$ is the value for which the characteristic curve goes to infinity:
\begin{equation}
r_{\rm min}(r,E_i)=r\left[1+r^2\frac{(1+C_a)}{C_bE_i}\right]^{-\frac{1}{1+C_a}}. 
\end{equation}
\section{Analytical estimate of the $p\gamma$ cooling rate with photons from electron synchrotron as targets }\label{app.pgamma}
In this appendix we estimate the cooling rate $t_{p\gamma}^{-1}$ in the case that the target photon density the synchrotron emission of electrons given by Eq.(\ref{nph}). If synchrotron cooling dominates for electrons, as is the case for the magnetic field values adopted, we can approximate the photon density by supposing that the same power injected in electrons is radiated. Since the corresponding electron distribution is $N_e\propto E_e^{-3}$ for a simplified injection of electrons $Q_e\sim K_e E_e^{-2}$, with 
$$ K_e=\frac{L_e}{4\pi \Gamma \Delta V_{b} \log\frac{E_{e,\rm max}}{E_{e,\rm min}}}$$ and $$E_{e,\rm max}=m_e c^2\sqrt{\frac{6\pi\,e\,\eta}{\sigma_{\rm T}B_{\rm acc}}},$$ 
 considering that the synchrotron emission is concentrated in the energy range given by $$E_{\rm ph}^{\rm min(max)}=\frac{\sqrt{6}heB_{\rm acc}}{4\pi m_e c}\left(\frac{E_{\rm min(max)}}{m_ec^2}\right)^2$$ leads to an emissivity $Q_{e,\rm syn}\approx \frac{K_e}{2} E_{\rm ph}^{-2}$. 
The density of such photons is, then: 
\begin{eqnarray}
 n_{\rm ph}=4\pi Q_{e,{\rm syn}}\frac{R_{j}}{c}\approx \frac{2K_e\pi R_j}{cE_{\rm ph}^{2}},
\end{eqnarray}
in units of [${\rm energy^{-1} length^{-3}}$].
In order to estimate the $t^{-1}_{p\gamma}$, we apply the approximation for the cross section given by Atoyan \& Dermer (2003) \cite{atoyandermer2003}, i.e.,
\begin{eqnarray}
\sigma_{p\gamma}(E_{\rm r})=\left\{\begin{array}{cc}
0  & \ \ \  {\rm for \ }  \  E_r< 0.2{\, \rm GeV} \\
340 \, \mu{\rm barn} & \ \ \ \ {\rm for \ } 0.2{\, \rm GeV} <E_r< 0.5{\, \rm GeV} \\
120 \, \mu{\rm barn} & {\rm for \ } E_r \ \geq 0.5{\, \rm GeV}, 
\end{array}\right.
\end{eqnarray}
where the low energy range corresponds to the single pion ($p+\gamma \rightarrow \pi^+ n$) with an inelasticity $K_1=0.2$, and for higher energies the multipion channel dominates ($p+\gamma \rightarrow p+ \pi^+ + \pi^- + \pi^0$) with $K_2=0.6$. In the case of the single-pion channel,
\begin{eqnarray}
 t^{-1}_{p\gamma,1}(\gamma_p)=K_1\int_{\frac{e_{\rm th}}{2\gamma_p}}^\infty dE_{\rm ph}\frac{c n_{\rm ph}(E_{\rm ph})}{2\gamma_p^2E_{\rm ph}^2}\int_{E_{\rm th}}^{2E_{\rm ph}\gamma_p}dE_r\sigma_1 E_r H(E_2 -E_r ) \nonumber 
\end{eqnarray} 
\begin{multline}
 t^{-1}_{p\gamma,1}(\gamma_p)\approx
\frac{ K_e K_1\sigma_1 R_j \pi }{2 \gamma_p^2}\int_{\frac{E_1}{2\gamma_p}}^{\frac{E_2}{2\gamma_p}} dE_{\rm ph}{E_{\rm ph}^{-4}} \left( 4 E_{\rm ph}^2\gamma_p^2-E_1^2 \right) +\\ \frac{K_e R_j\pi(E_2^2-E_1^2)}{2\gamma_p^2}\int_{\frac{E_2}{2\gamma_p}}^{E_{\rm ph,max}}dE_{\rm ph}{E_{\rm ph}^{-4}}\nonumber 
\end{multline}

\begin{eqnarray}
 t^{-1}_{p\gamma,1}(\gamma_p)&\approx& \frac{ 8 K_e K_1 \sigma_1 R_j  \pi (E_2-E_1)\gamma_p}{3 E_1E_2} \nonumber \\ &=& 1.1\times 10^{-8}{\rm s^{-1}}\left(\frac{E_p}{\rm GeV}\right) \left(\frac{L_e}{L_i}\right)\left(\frac{\Gamma}{1.67}\right)^{-1}. \label{omega1est}
\end{eqnarray}
For the multipion channel, we find,
\begin{eqnarray}
 t^{-1}_{\rm p\gamma,2}(\gamma_p)&=&\int_{\frac{E_{2}}{2\gamma_p}}^\infty dE_{\rm ph}\frac{c n_{\rm ph}(E_{\rm ph})}{2\gamma_p^2E_{\rm ph}^2} \int_{E_{2}}^{2E_{\rm ph}\gamma_p}dE_r\sigma_2 E_r \nonumber\\
 &=&\int_{\frac{E_2}{2\gamma_p}}^{E_{\rm ph,max}} dE_{\rm ph}\frac{c n_{\rm ph}(E_{\rm ph})}{2\gamma_p^2E_{\rm ph}^2} \frac{\sigma_2}{2} \left( 4 E_{\rm ph}^2\gamma_p^2-E_2^2 \right)\nonumber \\
&\approx&
\frac{ 8 K_e\sigma_2 R_j \pi \gamma_p}{3 E_2} \nonumber \\ &\approx&8\times 10^{-9}{\rm s^{-1}}\left(\frac{E_p}{\rm GeV}\right) \left(\frac{L_e}{5\times 10^{39}{\rm erg\,s^{-1}}}\right)\left(\frac{\Gamma}{1.67}\right)^{-1}. \nonumber
\end{eqnarray}
Hence, the total interaction rate can be approximated by
\begin{eqnarray}
t^{-1}_{p\gamma}(E_p)\approx 3\times 10^{-8}{\rm s^{-1}}\left(\frac{E_p}{\rm GeV}\right) \left(\frac{L_e}{L_i}\right)\left(\frac{\Gamma}{1.67}\right)^{-1},
\end{eqnarray}
and this matches the result shown in Fig.\ref{fig:pecool} for $E_p \lesssim 10^{8}$, while the exact result flattens at higher energies because the target photon density used in that case is corrected by synchrotron self-absorption.

\subsection{Estimation of neutrino emissivity}

In the case of a constant escape rate $t_{\rm esc}^{-1}= c/(\Gamma\Delta z)$, the distribution of protons at the base is roughly given by
$$
N_p(E_p)\approx  \frac{L_p  \Delta z_{\rm b}}{4\pi c \Delta V_{\rm b} \log\left(\frac{E_{p,{\rm max}}}{2m_p c^2}\right)} E_p^{-2},
$$
with $E_{p,\rm max}\approx 3\times 10^7{\rm GeV}$.
The emissivity of pions produced by $p\gamma$ interactions can be approximated using the collision frequency $\omega_{p\gamma}=t_{p\gamma}^{-1}/K_{p\gamma}$.
In turn, following Ref. \citep{atoyandermer2003}, the neutrino emissivity can be obtained approximately  by supposing that the pion energy is equally distributed among the four final decay products (after the muon decay). 
For the $\pi^+$ decay of the single pion channel, considering that the pion production takes place half of the times as compared to the neutron production, this leads to: 
\begin{eqnarray}
Q_{\nu_e}^{(1)}(E_\nu)=Q_{\bar{\nu}_\mu}^{(1)}(E_\nu)=Q_{{\nu}_\mu}^{(1)}(E_\nu)\approx 10 N_p(20 E_\nu){\omega_1(20E_\nu)}. \nonumber
\end{eqnarray}
For the decays the $\pi^+$ decay in the multipion channel, we have:
\begin{eqnarray}
Q_{\nu_e}^{(1)}(E_\nu)=Q_{\bar{\nu}_\mu}^{(1)}(E_\nu)=Q_{{\nu}_\mu}^{(1)}(E_\nu)\approx 20 N_p(20 E_\nu){\omega_2(20E_\nu)},\nonumber
\end{eqnarray}
 and for the $\pi^-$ decay of the multipion channel:
\begin{eqnarray}
Q_{{\nu}_\mu}^{(2)}(E_\nu)=Q_{\bar{\nu}_\mu}^{(2)}(E_\nu) =Q_{\bar{\nu}_e}^{(2)}(E_\nu) \approx 20 N_p(20 E_\nu){\omega_2(20E_\nu)}.\nonumber
\end{eqnarray}

The total emissivities for $\nu_\mu+\bar{\nu}_\mu$ and $\nu_e+\bar{\nu}_e$ are then:
\begin{eqnarray}
Q_{{\nu_\mu}+{\bar{\nu}_\mu}}(E_\nu)&=&2Q_{{\nu_e}+{\bar{\nu}_e}}(E_\nu)\approx 20 \omega_1(20 E_\nu)N_p(20 E_\nu)\nonumber \\&+& 80 \omega_2(20E_\nu)N_p(E_\nu) \nonumber \\
&\approx& 150 \, {\rm GeV^{-1}cm^{-3}sr^{-1}s^{-1}}\left(\frac{E_\nu}{\rm GeV} \right)^{-1} \nonumber \\ &\times&\left(\frac{L_p}{L_i}\right)  \left(\frac{L_e}{L_i}\right)\left(\frac{\Gamma}{1.67}\right)^{-1} \label{Qnumu_nue_esti}.
\end{eqnarray} 
In this case, these simplified expressions yield results within a factor $\sim 0.5$ of the obtained with the more accurate treatment of Ref. \cite{hummer2010}, but can still be useful as order of magnitude estimates.

\section*{Acknowledgments}

  AMC and MMR are supported by grants PIP 0046 (CONICET) and 15/E870EXA912/18 (Universidad Nacional de Mar del Plata). GER is supported by grant PIP 0338 (CONICET), PICT 2017-2865 (ANPCyT), and by the Ministerio de Econom\'{\i}a y Competitividad (MINECO) under grant AYA2016-76012-C3-1-P and PID 2019-105510GB-C31.

\end{document}